\documentclass[aps,pre,twocolumn,superscriptaddress,showpacs]{revtex4-2}
\usepackage{color}
\usepackage{amssymb}
\usepackage{amsmath}
\usepackage{bm}
\usepackage{url}
\usepackage{graphicx}

\begin{document}
\title{Effective models and predictability of chaotic multiscale systems via machine learning}

\author{Francesco Borra}
\thanks{Corresponding author}
\email{francesco.borra@uniroma1.it}
\affiliation{Dipartimento di Fisica, Universit\`a ``Sapienza''  Piazzale A. Moro 5, I-00185 Rome, Italy}

\author{Angelo Vulpiani}
\affiliation{Dipartimento di Fisica, Universit\`a ``Sapienza''  Piazzale A. Moro 5, I-00185 Rome, Italy}

\author{Massimo Cencini}
\thanks{Corresponding author}
\email{massimo.cencini@cnr.it}
\affiliation{Istituto dei Sistemi Complessi, CNR, via dei Taurini 19, I-00185 Rome, Italy}

\begin{abstract}
  Understanding and modeling the dynamics of
    multiscale systems is a problem of considerable interest both for
    theory and applications. For unavoidable practical reasons, in
    multiscale systems, there is the need to eliminate from the description
    the fast/small-scale degrees of freedom and thus build effective models
    for only the slow/large-scale degrees of freedom.  When there is a
    wide scale separation between the degrees of freedom, asymptotic
    techniques, such as the adiabatic approximation,
    can be used for devising such effective models, while away from this limit
    there exist no systematic techniques. Here, we scrutinize the use of machine learning, based
  on reservoir computing, to build data-driven effective models of
  multiscale chaotic systems. We show that, for a wide scale
  separation, machine learning generates effective models akin to
  those obtained using multiscale asymptotic techniques and,
  remarkably, remains effective in predictability also when the scale
  separation is reduced. We also show that predictability can be
  improved by hybridizing the reservoir with an imperfect model.
\end{abstract}


\maketitle

\section{Introduction\label{sec:intro}}
Machine learning techniques are impacting science at an impressive
pace from robotics~\cite{argall2009survey} to genetics~\cite{libbrecht2015machine}, medicine~\cite{he2019practical}, and
physics~\cite{carleo2019machine}. In physics, reservoir
computing \cite{verstraeten2007experimental,schrauwen2007overview}, 
based on echo-state neural networks
\cite{jaeger2001echo,JaegerReview,jaegerScience}, is gathering much
attention for model-free, data-driven predictions of chaotic evolutions
\cite{ottLyapunov,ottPRL,ottAttractor,vlachas2018data,nakai2018machine}.
Here, we scrutinize the use of reservoir computing  to
build effective models for predicting the slow degrees of freedom of
multiscale chaotic systems. We also consider hybrid reservoirs,
blending data with predictions based on an imperfect model
\cite{ottHybrid} (see also Ref.~\cite{wikner2020combining}).

Multiscale chaotic systems represent a challenge to both theory and
applications. For instance, turbulence can easily span over 4/6
decades in temporal/spatial scales \cite{warhaft2002turbulence}, while
climate time scales range from hours of atmosphere variability to
thousands years of deep ocean currents
\cite{peixoto1992physics,pedlosky2013geophysical}.  These huge ranges
of scales stymie direct numerical approaches making modeling of fast
degrees of freedom mandatory, being slow ones usually the most
interesting to predict. In principle, the latter are easier to
predict: the maximal Lyapunov exponent (of the order of the inverse of
the fastest time scale) controls the early dynamics of very small
perturbations appertaining to the fast degrees of freedom that
saturate with time, letting the perturbations on the slow degrees of
freedom to grow at a slower rate controlled by the typically weaker
nonlinear instabilities
\cite{lorenz1996predictability,aurell1996growth,cencini2013finite}.
However, owing to nonlinearity, fast degrees of freedom depend on, and
in turn, impact on the slower ones. Consequently, improper modeling
the former severely hampers the predictability of the
latter~\cite{boffetta2000predictability}.

We focus here on a simplified setting with only two time scales,
i.e. on systems of the form:
\begin{equation}
  \begin{aligned}
  \dot{\bm X}&=\frac{1}{\tau_s}\bm F_s(\bm X,\bm x)\\
  \dot{\bm x}&=\frac{1}{\tau_f}\bm F_f(\bm x,\bm X)\,,
\end{aligned}
  \label{eq:1}
\end{equation}
where $\bm X$ and $\bm x$ represent the slow and fast degrees of
freedom, respectively. The time scale separation between them is
controlled by $c\!=\!\tau_s/\tau_f$. The goal is to build an effective
model for the slow variables, $\dot{\bm X}\!=\!\!\bm
F_{\mathrm{eff}}(\bm X)$, to predict their evolution.  When the fast
variables are much faster than the slow ones ($c \gg 1$), multiscale
techniques \cite{sanders2007averaging,pavliotis2008multiscale}
can be used to build effective models.  Aside from
such limit, systematic methods for deriving effective models are
typically unavailable.

In this article, we show that reservoir computers  trained on
time series of the slow degrees of freedom can be optimized to build
(model-free data-driven) effective models able to predict the slow dynamics. Provided the
reservoir dimensionality is high enough, the method works both when
the scale separation is large, red basically recovering the results of standard multiscale methods, such as the adiabatic approximation, and when it not so large. Moreover, we show that even an imperfect knowledge of the slow dynamics can be used to improve predictability, 
also for smaller reservoirs. 

The material is organized as follows. In Sec.~\ref{sec:implementation}
we present the reservoir computing approach for predicting chaotic
systems, moreover we provide the basics of its implementation also
considering the case in which an imperfect model is available (hybrid
implementation). In Sec.~\ref{sec:result} we present
  the main results obtained with a specific multiscale system.
Section~\ref{sec:end} is devoted to discussions and perspectives. In
Appendix~\ref{app:implementation} we give further details on
implementation, including the choice of
hyperparameters. Appendix~\ref{app:model} presents the adiabatic
approximation for the multiscale system here considered. In
Appendix~\ref{app:hybrid} we discuss and compare different hybrid
schemes.

\section{Reservoir computing for chaotic systems and its implementation \label{sec:implementation}}

Reservoir computing
\cite{verstraeten2007experimental,schrauwen2007overview} is a
brain inspired approach based on a recurrent neural network (RNN), the
reservoir (R) -- i.e. an auxiliary high dimensional nonlinear
dynamical system naturally suited to deal with time sequences--,
(usually) linearly coupled to a time dependent lower dimensional input
(I), to produce an output (O). To make O optimized for approximating
some desired dynamical observable, the network must be
trained. Reservoir computing implementation avoids backpropagation
\cite{werbos1990backpropagation} by only training the output layer,
while R-to-R and I-to-R connections are quenched random
variables. Remarkably, the reservoir computing approach allows for
fast hardware implementations with a variety of nonlinear systems
\cite{larger2017high,tanaka2019recent}. Choosing the output as a
linear projection of functions of the R-state, the optimization can be
rapidly achieved via linear regression. The method works provided
R-to-R connections are designed to force the R-state to only depend on
the recent past history of the input signal, fading the memory of the
initial state.

\subsection{Predicting chaotic systems with reservoir computing} 
\begin{figure}[t!]
\centering
\includegraphics[width=1\columnwidth]{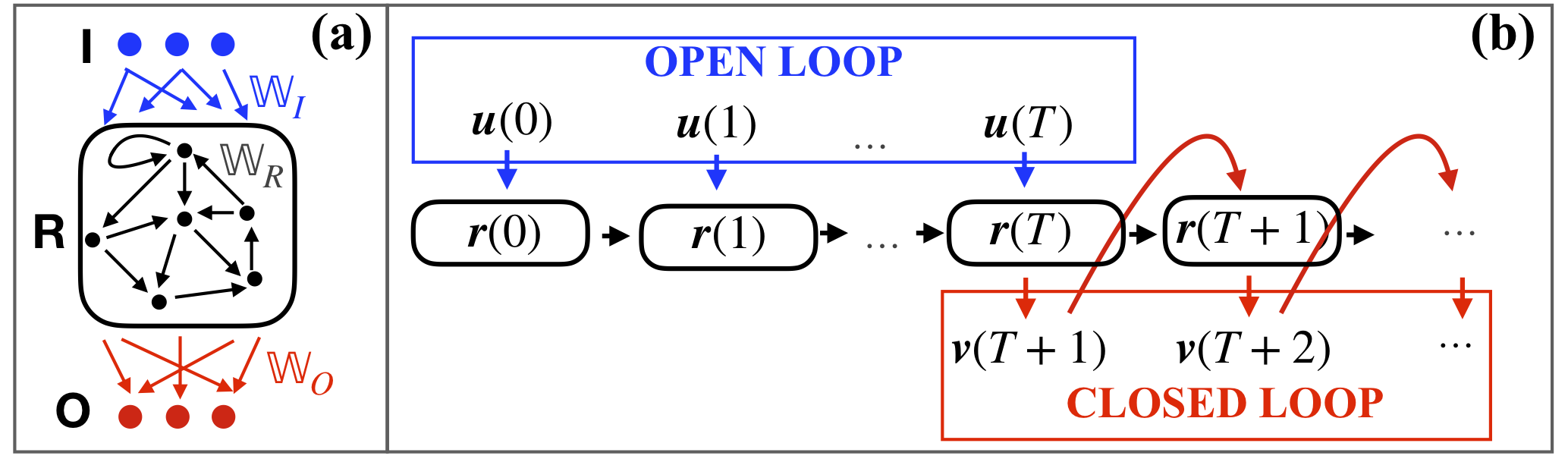}
\caption{(Color online) Sketch of  reservoir computing: (a) the components and their connections; (b) the two modes of operation: open loop for synchronizing the reservoir to the input and for training, closed loop for prediction. \label{fig:1}}
\vspace{-0.5truecm}
\end{figure}

When considering a chaotic dynamical system with state $\bm s(t)=(\bm
X(t),\bm x(t))$, with reference to Eqs.~(\ref{eq:1}), the input signal
$\bm u(t)\in \mathrm{I\!R}^{D_{I}}$ is typically a subset of the state
observables, $\bm u(t)\!=\!\bm h(\bm s(t))$. For
  instance, in the following we consider functions of the slow
variables, $\bm X$, only. When the dimensionality, $D_R$, of the
reservoir is large enough and the R-to-R connections are suitable
chosen, its state, $\bm r(t)\in \mathrm{I\!R}^{D_{R}}$, becomes a
representation -- an echo -- of the input state $\bm s(t)$
\cite{jaegerScience,schrauwen2007overview,ottAttractor}, via a
mechanism similar to generalized synchronization
\cite{ottAttractor,pikovsky2003synchronization}. In this
configuration, dubbed open loop \cite{rivkind2017local}
(Fig.~\ref{fig:1}a), the RNN is driven by the input and, loosely
speaking, synchronizes with it.  When this is achieved, the output,
$\bm v(t)\in \mathrm{I\!R}^{D_{O}}$ can be trained (optimized) to fit
a desired function of $\bm s(t)$, for instance, to predict the next
outcome of the observable, i.e. $\bm v(t+\Delta t) =\bm u(t+\Delta
t)$. After training, we can close the loop by
  feeding the output as a new input to R (Fig.~\ref{fig:1}b), thus
  obtaining an effective model for predicting the time sequence.  For
the closed loop mode to constitute an effective (neural) model of the
dynamics of interest, we ask the network to work for arbitrary initial
conditions, i.e. not only right after the training: a property dubbed
\textit{reusability} in Ref. ~\cite{ottAttractor}.
For this purpose, when starting from a random reservoir state, a short
synchronization period in open loop is needed before closing the loop.
The method to work requires some stability property which cannot, in
general, be granted in the closed loop configuration
\cite{rivkind2017local}.


\subsection{Implementation\label{sec:RC}} 
Reservoir neurons can be implemented in different ways
\cite{verstraeten2007experimental}, we use echo state neural network
\cite{jaegerScience}, mostly following
\cite{ottLyapunov,ottPRL,ottAttractor}. Here, we assume $D_{R}\!\gg\!
D_{I}\!=\!D_{O}$ and the input to be sampled at discrete time
intervals $\Delta t$. Both assumptions are not
  restrictive, for instance in the hybrid implementation below we will
  use $D_O\neq D_I$ and the extension to continuous time is
  straightforward \cite{verstraeten2007experimental}. The reservoir
is built via a sparse (low degree, $d$), random graph
  represented via a $D_R\times D_R$ connectivity matrix
$\mathbb{W}_R$, with the non zero entries uniformly
  distributed in $[-1,\;1]$, scaled to have a specific spectral
radius $\rho=\max\{|\mu_i|\}$ with $\mu_i$ being the
  matrix eigenvalues. The request $\rho<1$ is sufficient, though not
strictly necessary \cite{jiang2019model}, to ensure the echo state
property \cite{jaeger2001echo,JaegerReview} in open loop, namely the
synchronization of $\bm r(t)$ with $\bm s(t)$. We distinguish training
and prediction. Training is done in open loop mode using an input
trajectory $\bm u(t)$ with $t\in[-T_s,T_t]$, where $T_t$
  is the training input sequence length, and $T_s$ is the length of
  initial transient to let the $\bm r(t)$, randomly initialized at
  $t=-T_s$, to synchronize with the system dynamics. After being
scaled to be zero mean and unit standard deviation, the input is
linearly coupled to the reservoir nodes via a $D_R\times D_{I}$ matrix
$\mathbb{W}_{I}$, with the non zero entries taken as
  random variables uniformly distributed in $[-\sigma,\;\sigma]$. In
open loop mode the network state $\bm r(t)$ is updated
  as
\begin{equation}
\bm r(t+\Delta t)=\tanh[\mathbb{W}_R\bm r(t)+\mathbb{W}_{I} \bm u(t)]\,.
\label{eq:2}
\end{equation}
In the above expression the $\tanh$ is applied
element wise, and can be replaced with other nonlinearities.  The
output is computed as $\bm v(t+\Delta t)=\mathbb{W}_O\bm
r^\star(t+\Delta t)$ with the $D_R\times D_{O}$ matrix $\mathbb{W}_O$
obtained via linear regression by imposing
$\mathbb{W}_O=\arg\min_{\mathbb{W}} \{\sum_{0\leq t\leq T_t}||\bm
v(t)-\bm u(t)||^2+\alpha \mathrm{Tr}[\mathbb{W}\mathbb{W}^T] \}$, to
ensure the output to be the best predictor of the next input
observable. The term proportional to $\alpha$ is a regularization,
while $\bm r^\star$ is a function of the reservoir state. Here, we
take $r^*_i(t)=r_i(t)$ if $i$ is odd and $r^*_i(t)=r^2_i(t)$
otherwise \cite{Note1}.  Once $\mathbb{W}_O$ is
determined, we switch to prediction mode. Given a short sequence of
measurements, in open loop, we can synchronize the reservoir with the
dynamics (\ref{eq:2}), and then close the loop letting $\bm
u(t)\leftarrow \bm v(t)=\mathbb{W}_O\bm r^\star(t)$ in
Eq.~(\ref{eq:2}).  This way Eq.~(\ref{eq:2}) becomes a fully data-driven effective model for the time signal to be predicted. The
resulting model, and thus its performances, will
  implicitly depend both on the hyperparameters ($d,\rho$ and
  $\sigma$) defining the RNN structure and the I-to-R connections
  and on the length of the training trajectory ($T_t$). The choices of
  these hyperparameters are discussed in
  Appendix~\ref{app:implementation}.


\subsection{Hybrid implementation\label{sec:hybrid}}
So far we assumed no prior knowledge of the dynamical system that
generated the input.  If we have an imperfect model for approximately
predicting the next outcome of the observables $\bm u(t)$, we can
include such information in a hybrid scheme by slightly changing the
input and/or output scheme to exploit this extra
  knowledge \cite{ottHybrid,wikner2020combining}.  The idea of
  blending machine learning algorithms with physics informed model is
  quite general and it has been exploited also with methods different
  from reservoir computing, see
  e.g. Refs.~\cite{milano2002neural,wan2018data,weymouth2013physics}.

\begin{figure*}[t!]
\centering
\includegraphics[width=1\textwidth]{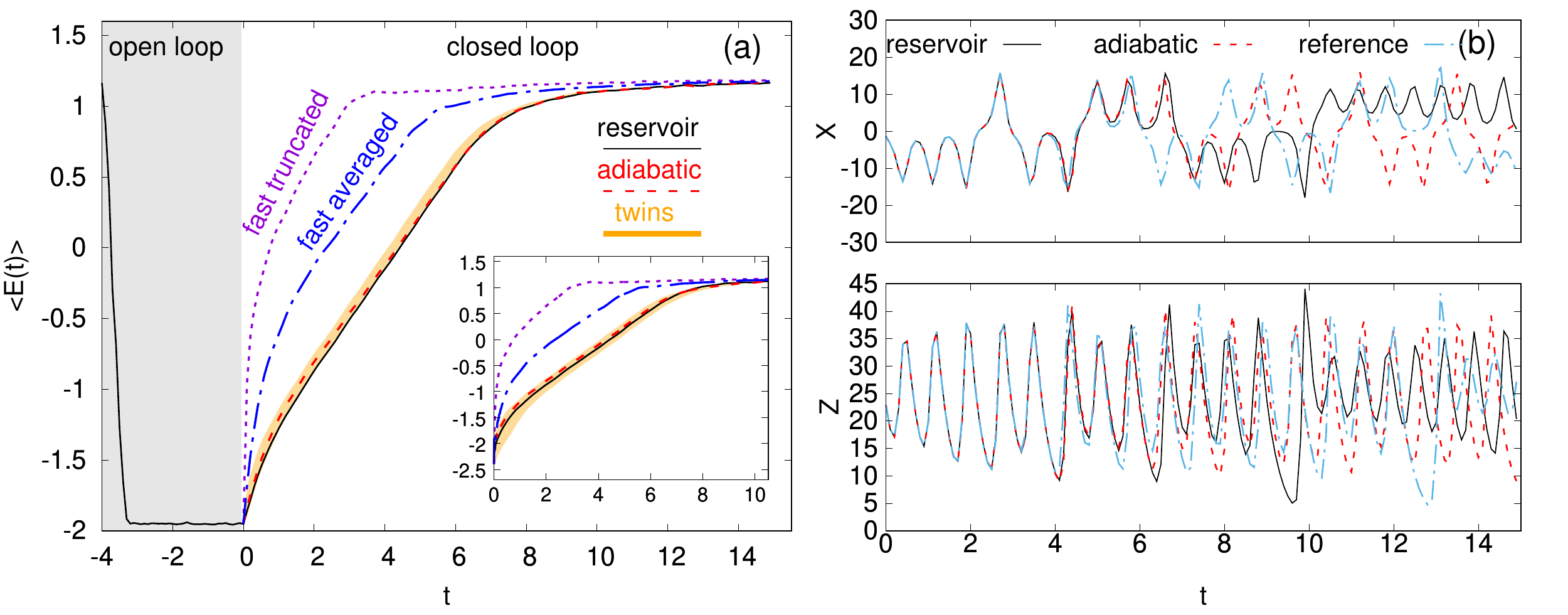}
\vspace{-0.7truecm}
\caption{(Color online) Prediction error growth  for a single realization of
  a network of $D_R\!=\!500$ neurons.  (a) Average (over $10^4$ initial
  conditions) ($\log_{10}$)error $\langle E(t)\rangle$ vs time during
  synchronization (open loop, gray region) and prediction
  (closed loop) for $c\!=\!10$ and $\Delta t\!=\!0.1$: the yellow shaded area
  circumscribes the twin and random twin
  model predictions (see text); reservoir computer prediction (solid, black
  curve) compared with that of the truncated model (purple, dotted curve),
  of the model fast variables replaced by their average (blue, dash
  dotted curve) and model (\ref{eq:improved}) (red, dashed curve). The
  inset shows the same (closed loop only) for $\Delta
  t\!=\!0.01$. (b) An instance of a prediction experiment, showing the reference (dash dotted, light blue curves) evolution of the $X$ (top) and $Z$ (bottom) variables of the coupled Lorenz model (\ref{eq:lorenzslow}) together with the prediction obtained via the reservoir (black, solid curve) and the adiabatic model (dashed, red curve).
  For details on hyperparameters see Appendix~\ref{app:hyper}.
 \label{fig:2}}
\end{figure*}

Let $\wp[\bm u(t)]=\hat{\bm u}(t+\Delta t)$ be the estimated next
outcome of the observable $\bm u(t)$ according to our imperfect
model. The idea is to supply such information in the input by
replacing $\bm u(t)$ with the column vector $(\bm u(t),\wp[\bm
  u(t)])^T$, thus doubling the dimensionality of the input matrix. For
the output we proceed as before. The whole scheme is thus as the above
one with the only difference that $\mathbb{W}_O$ is now a $D_R\times
D_I/2$. The switch to the prediction mode is then obtained using
$(\mathbb{W}_O\bm r^\star (t),\wp[\mathbb{W}_O\bm r^\star (t)])^T$ as
input in Eq.~(\ref{eq:2}).

It is worth noticing that other hybrid schemes are possible, e.g. in
Ref.~\cite{ottHybrid} the output has the form $\bm v(t+\Delta
t)=\mathbb{W}_O(\bm r^\star(t),\wp[\bm u(t)])^T$, namely a
combination of the prediction based on the network and on the physical
model. In Appendix~\ref{app:hybrid} we comment further on our choice,
and we compare it with the scheme proposed in Ref.~\cite{ottHybrid}.

\section{Results for  a two time scales system \label{sec:result}}
We now consider the model introduced in Ref.~\cite{boffetta1998extension} as a caricature for the interaction of the (fast) atmosphere and the (slow) ocean. It consists of two Lorenz systems coupled as follows:
\begin{eqnarray}
&&\left\{
  \begin{array}{l}
 \dot{X}=a(Y-X)
 \\ \dot{Y}=R_sX-ZX-Y-\epsilon_sxy
 \\ \dot{Z}=XY-bZ
  \end{array}
    \right.
  \label{eq:lorenzslow}\\
 & &\left\{
  \begin{array}{ll}
 \dot{x}=ca(y-x)\\
 \dot{y}=c(R_fx-zx-y)+\epsilon_fYx\\
 \dot{z}=c(xy-bz)\,,
  \end{array}
  \right.
\label{eq:lorenzfast}
\end{eqnarray}  
  where Eqs. (\ref{eq:lorenzslow}) and Eqs. (\ref{eq:lorenzfast}) describe the
  evolution of the slow and fast variables, respectively.  We fix the
parameters as in Ref.~\cite{boffetta1998extension}: $a=10$, $b\!=\!8/3$,
$R_s\!=\!28$, $R_f\!=\!45$, $\epsilon_s\!=\!10^{-2}$ and
$\epsilon_f\!=\!10$, while for the time scale separation parameter,
$c$, we use $c\!=\!10$ (as in Ref.~\cite{boffetta1998extension}) and
$c\!=\!3$. The former corresponds to a scale separation such that the
adiabatic approximation already provides good results (see below).
Moreover, for $c\!=\!10$, the error growth on the slow variables is
characterized by two exponential regimes \cite{boffetta1998extension}:
the former with rate given by the Lyapunov exponent of the full system
$\lambda_f\!\approx\! 11.5$, and the latter by $\lambda_s\!\approx\!
0.85$, controlled by the fast and slow instabilities,
respectively. This decomposition can be made more rigorous as shown in
Ref.~\cite{ginelli} for a closely related model.

We test the reservoir computing approach inputting the slow variables,
i.e. $\bm u(t)\!=\!(X(t),Y(t),Z(t))$.  In open loop, we let the
reservoir to synchronize with the input, subsequently we perform the
training and optimize $\mathbb{W}_O$ as explained earlier. Then, to
test the prediction performance we consider $10^4$ initial conditions,
for each of which, we feed the slow variables to the network in open
loop and record, from $t\!=\!-T_s$ to $t\!=\!0$, the one step
($\log_{10}$)error $E(t)\!=\!\log_{10}\| {\bm v}(t)-{\bm u}(t)\|$,
${\bm v}(t)$ being the one-step network forecast (output).

\begin{figure*}[t!]
\centering
\includegraphics[width=1\textwidth]{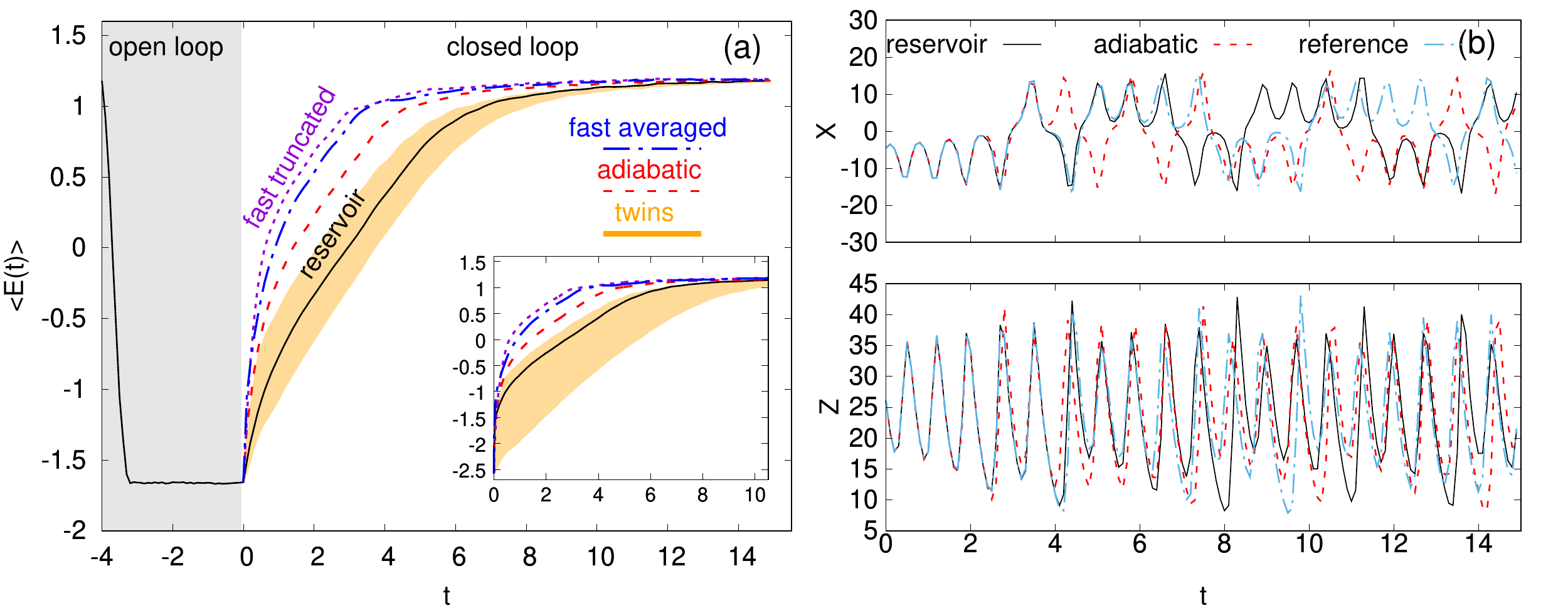}
\vspace{-0.7truecm}
\caption{(Color online) Same as Fig.~\ref{fig:2} for the case $c=3$.
 \label{fig:3}}
\end{figure*}

Initially, the average ($\log_{10}$)error $\langle
  E(t)\rangle$ decreases linearly as shown in the grey regions of
  Figs.~\ref{fig:2}a and \ref{fig:3}a, which is a visual proof of the
  echo state property. Then, it reaches a plateau - the
  synchronization error $E_S$ - which can be interpreted as the
  average smallest ($\log_{10}$)error on the initial condition and the
  one step error prediction.

At the end of the open loop, after synchronization, we switch to the
prediction (closed loop) configuration and compute the
($\log_{10}$)error growth between the network
  prediction and the reference trajectory. Moreover, we take the
output variables at the end of the open loop and use them
  as initial conditions for  other models (discussed below) which are used as a comparison.
  First, we consider the perfect model with an error on the
initial condition, i.e Eqs.~(\ref{eq:lorenzslow}) with the slow
variables set equal to the network-obtained values at $t=0$, i.e. at
the end of the open loop.  By construction, the network does not
forecast the fast variables, which are thus initialized either using
their exact values from the reference trajectory (twin model), which
is quite ``unfair'', or random values (random twin)
from the stationary measure on the fast attractor with fixed slow
variables. Then we consider increasingly refined effective models for
the slow degrees of freedom only: a ``truncated'' model, $\dot {\bm
  X}\!= \!\bm{F}_T(\bm X)$, obtained from Eqs.~(\ref{eq:lorenzslow})
by setting $\epsilon_s\!=\!0$; a model in which we replace the fast
variables in Eqs.~(\ref{eq:lorenzslow}) with their global average; 
the adiabatic model in which fast variables are averaged with fixed slow variables,
which amounts to replacing $\epsilon_s xy$ in the equation for
$\dot{Y}$ with (see Appendix~\ref{app:model} for
  details on the derivation):
\begin{equation}
  \epsilon_s \langle xy\rangle_{\bm X}\!=\!(1.07\!+\!0.26Y/c) \,\Theta(\!1.07\!+\!0.26Y/c)\,,
  \label{eq:improved}
\end{equation}
where $\Theta$ denotes the Heaviside step function.

 In Fig.~\ref{fig:2} we show the results of the
  comparison between the prediction obtained with the reservoir
  computing approach and the different models above described
  for $c=10$, with sampling time $\Delta t=0.1$ (and $\Delta t=0.01$ in
  the inset of Fig.~\ref{fig:2}a).  Figure~\ref{fig:2}a shows that
  eliminating the fast degrees of freedom (truncated model) or just
  considering their average effect leads to very poor predictions,
  while the prediction of the reservoir computer is comparable to that
  of the adiabatic model (\ref{eq:improved}), as qualitatively shown
  in Fig.~\ref{fig:2}b (whose top/bottom panels show the evolution of
  the slow variables $X$ and $Z$ for the reference trajectory and the
  predictions obtained via the reservoir and adiabatic model).
  Remarkably, the reservoir-based model seems to even
  slightly outperform the twin model.  A fact we
  understand as follows: by omitting fast components, one does not add
  fast decorrelating fluctuations to those intrinsic to the reference
  trajectories, thus reducing effective noise. Notice that the zero
  error on fast components of the twin model is rapidly pushed to its
  saturation value by the error on the slow variables. The sampling
  time $\Delta t\!=\!0.1$ is likely playing an important role during
  learning by acting as a low passing filter. Indeed the comparison
  with twin model slightly deteriorates for $\Delta t\!=\!0.01$ (see
  Fig.~\ref{fig:2}a inset).

Figure~\ref{fig:3} shows the results for
  $c\!=\!3$. Here, the poor scale separation spoils the effectiveness
  of the adiabatic model (\ref{eq:improved}) while the prediction
  obtained via the reservoir computing approach remains effective, as
  visually exemplified in Fig.~\ref{fig:3}b and quantified in
  Fig.~\ref{fig:3}a. Notice that, however, the network predictability
  deteriorates with respect to the previous case and the twin model
  does better, though the reservoir still outperforms the random twin
  model. This slight worsening is likely due to the fact that
  discarded variables are not fast enough to average themselves out,
  making the learning task harder. Nevertheless, the network remains
predictive.

\subsection{Which effective model the network has built?}
\begin{figure}[t!]
\centering
\includegraphics[width=1\columnwidth]{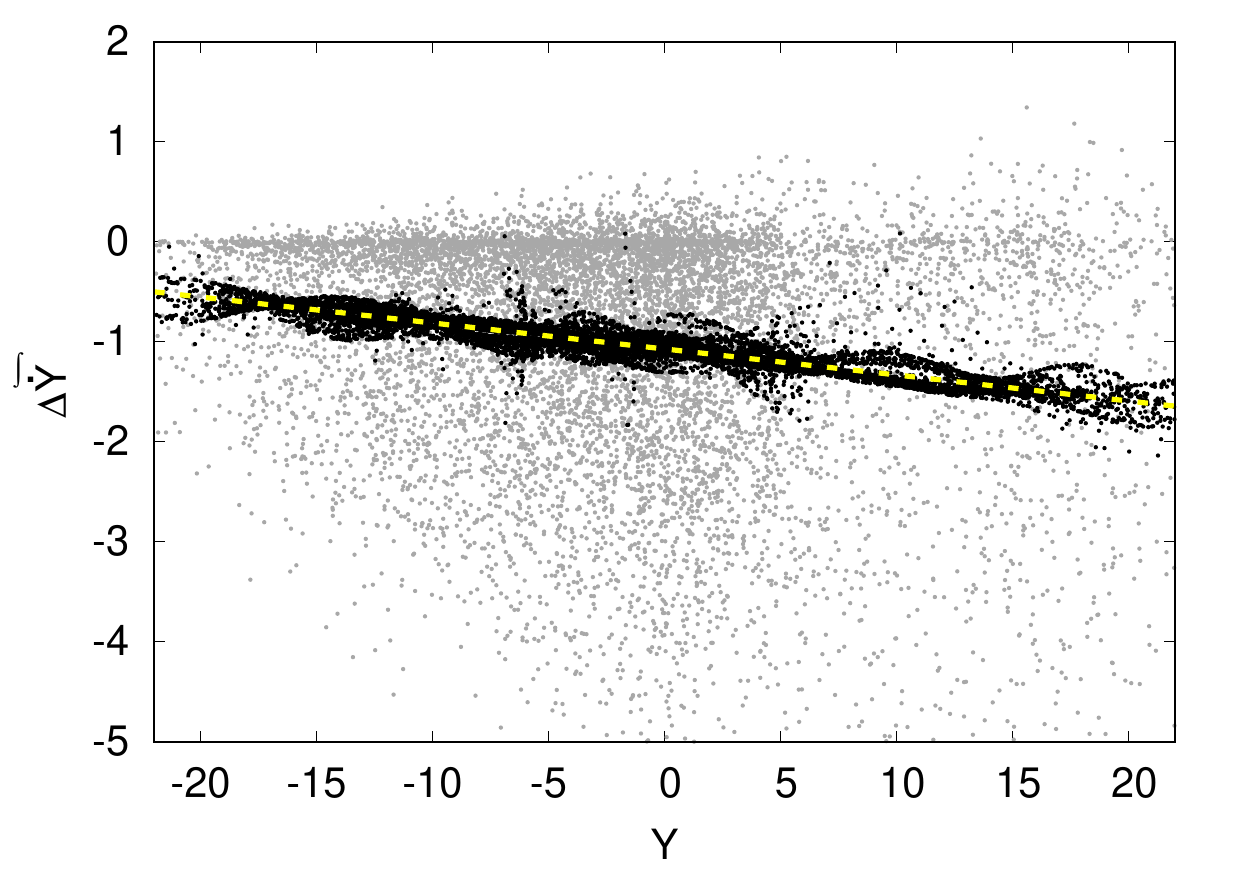}
\vspace{-0.7truecm}
\caption{(Color online) Residual derivatives
  (\ref{eq:dderiv}) vs $Y$ for $c\!=\!10$ and $\Delta
  t\!=\!0.01$, computed with the network (black dots), the multiscale
  model (\ref{eq:improved}) (yellow, dashed line), and the full
  dynamics (gray dots).
  For details on hyperparameters see Appendix~\ref{app:hyper}.
 \label{fig:4}}
\end{figure}

We now focus on the case $c\!=\!10$ and $\Delta t\!=\!0.01$, for which we can
gain some insights into how the network works by comparing it to the adiabatic
model~(\ref{eq:improved}). The sampling time is indeed small enough for
time differences to approximate derivatives. In Fig.~\ref{fig:4} we
demonstrate that the network in fact generates an effective model akin to the
adiabatic one (\ref{eq:improved}). Here we  show a surrogate of the
residual time derivative of $Y$, meaning that we removed the truncated model
derivative, as a function of $Y$:
\begin{equation}
  \Delta \widetilde{\dot{Y}}= \frac{Y(t+\Delta t)-Y(t)}{\Delta t}-\frac{Y_T(t+\Delta t)-Y(t)}{\Delta t}\,,
  \label{eq:dderiv}
\end{equation}
The expression in Eq.~(\ref{eq:dderiv}) provides a
  proxy for how the network has modeled the term $-\epsilon_s xy$ in
Eqs.~(\ref{eq:lorenzslow}).  The underlying idea is as follows. We let
the network evolve in closed loop, at time $t$ it takes as input the
forecasted slow variables $\bm
v(t)\!=\!(\hat{X}(t),\hat{Y}(t),\hat{Z}(t))$ and it outputs the next
step forecast $\bm v(t\!+\!\Delta t)\!=\!(\hat{X}(t\!+\!\Delta
t),\hat{Y}(t\!+\!\Delta t),\hat{Z}(t\!+\!\Delta t))$. We then use $\bm
v(t)$ as input to the truncated model, and evolve it for a time step
$\Delta t$ to obtain $(X_T(t+\Delta t),Y_T(t+\Delta t),Z_T(t+\Delta
t))$. Equation~(\ref{eq:dderiv}) is then used to infer how the network
models the coupling with the fast variables.  Evolving by one time
step $\bm v(t)$ using Eqs.~(\ref{eq:improved}) and then again
employing (\ref{eq:dderiv}) we, obviously, obtain the line
$-1.07-0.26Y$ (dashed in Fig.~\ref{fig:2}c).
The network residual
derivatives (black dots in Fig.~\ref{fig:4}) distribute on a narrow stripe
around that line. This means that the network, for wide scale separation,
performs an average conditioned on the values of the
  slow variables. For  $c=10$, such conditional average is equivalent
  to the adiabatic approximation (\ref{eq:improved}), as discussed in
Appendix~\ref{app:model}. For comparison, we
also show the residual derivatives (\ref{eq:dderiv})
computed with  the full model
(\ref{eq:lorenzslow}-\ref{eq:lorenzfast}) (gray dots), which display
a scattered distribution, best fitted by
Eq.~(\ref{eq:improved}). For $c=3$, while the
  adiabatic approximation is too rough, remarkably the network still
  performs well even though is more difficult to identify the model it
  has build, which will depend on the whole set of slow variables
 (see Appendix~\ref{app:model} for a further
    discussion).

\subsection{Predictability time and hybrid scheme} So far we focused on the predictability of a quite large network
($D_R\!=\!500$ as compared to the low dimensionality of
Eqs.~(\ref{eq:lorenzslow}-\ref{eq:lorenzfast})). How does the network
performances depend on the reservoir size $D_R$?

\begin{figure}[t!]
\centering
\includegraphics[width=1\columnwidth]{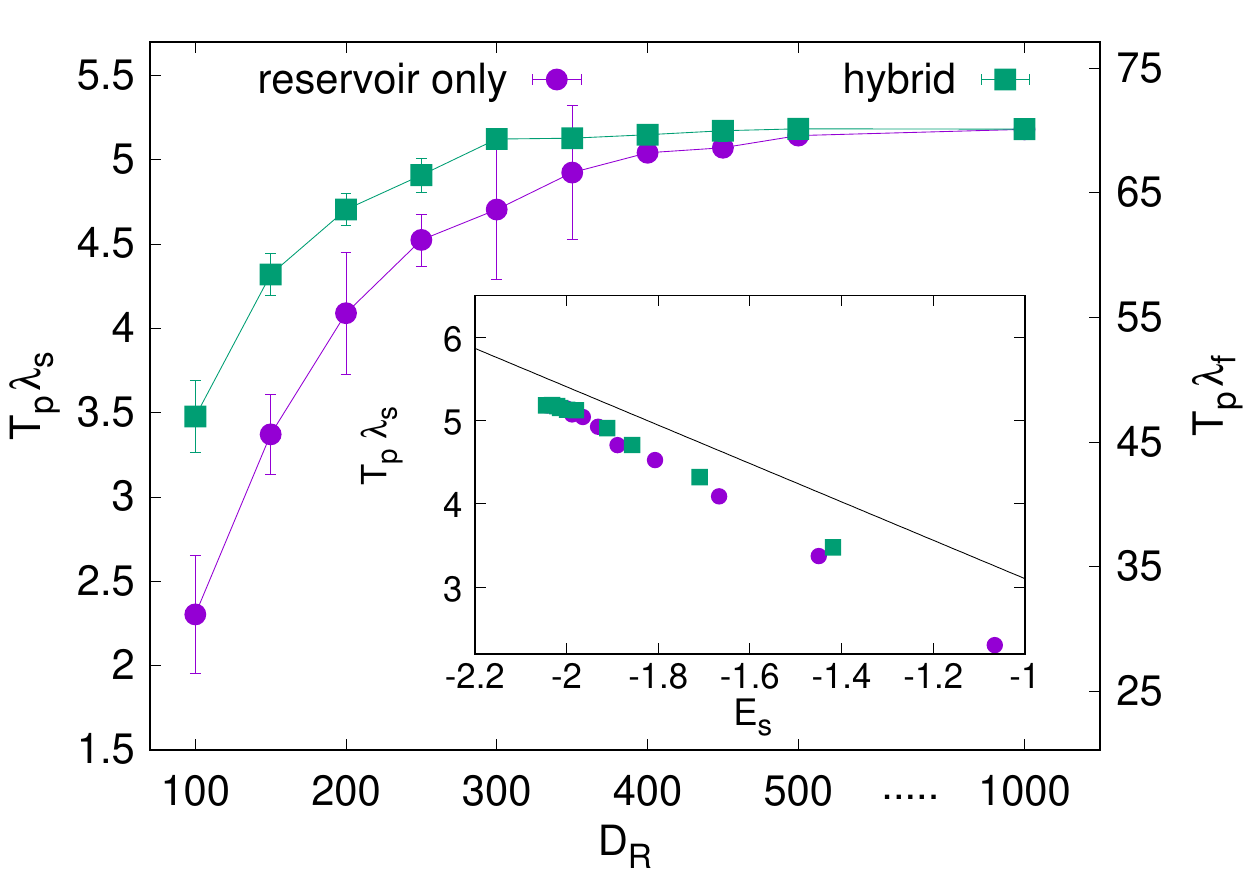}
\vspace{-0.7truecm}
\caption{(Color online) Average predictability time, $T_p$ normalized with the slow finite size Lyapunov exponent $\lambda_s$ (left scale) and with the (fast) maximal Lyapunov exponent $\lambda_f$ (right scale), versus
  reservoir size $D_R$ (hyperparameters for the hybrid implementation are the same of the reservoir only approach which are discussed in
  Appendix~\ref{app:hyper}) for reservoir only (purple circles) and hybrid
  scheme (green squares), system parameters $c\!=\!10$ and $\Delta
  t\!=\!0.1$. Error bars denote statistical standard deviation over 20
  independent network realizations, each sampling $10^3$ initial
  conditions.  Inset: $T_p\lambda_s$ vs synchronization average
  ($\log_{10}$)error $\langle E_S\rangle$. The slope of the black line
  is $-1$ corresponding to the  slow perturbation growth rate
  $\lambda_s$.\label{fig:5}}
\end{figure}

\begin{figure}[t!]
\centering
\includegraphics[width=1\columnwidth]{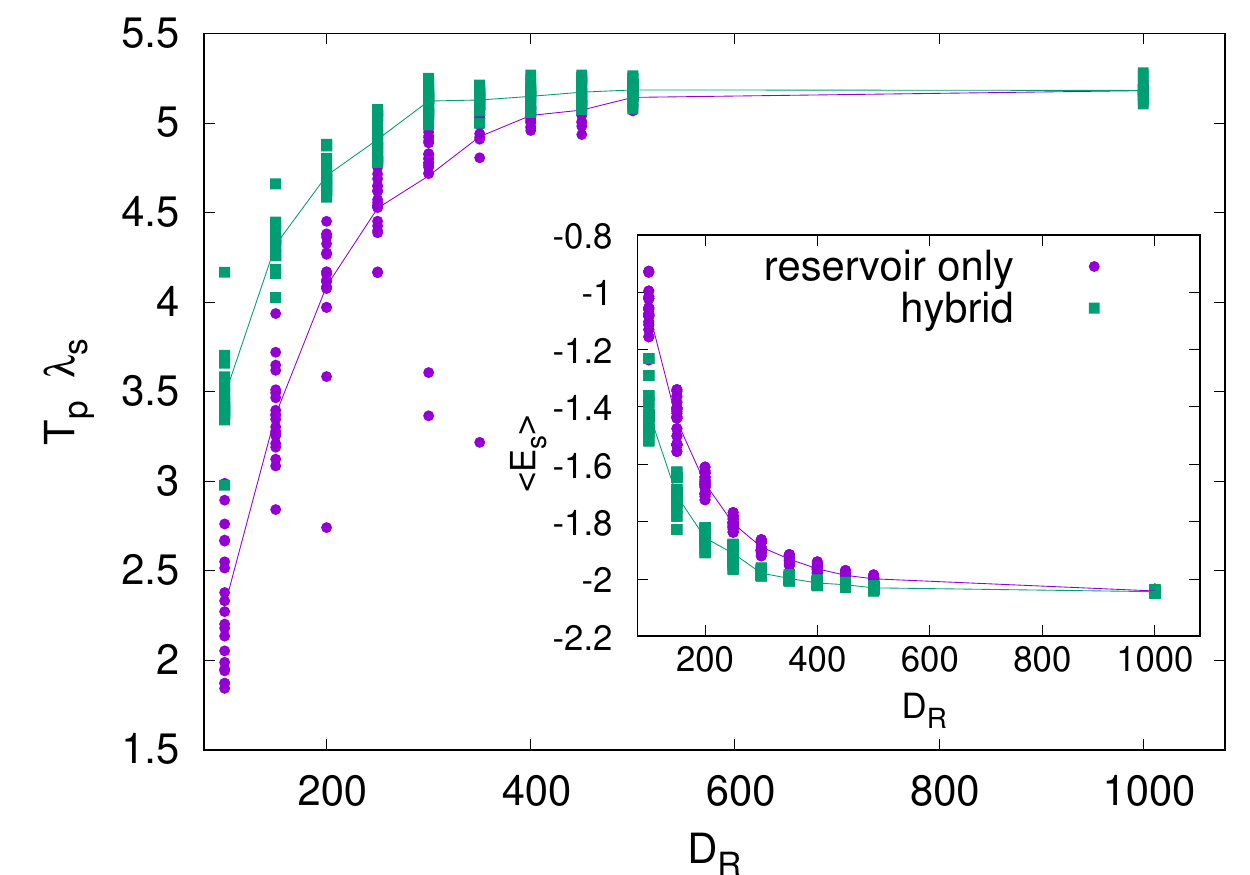}
\caption{(Color online) Nondimensional predictability time
  $T_p\lambda_s$ of $20$  network realizations (each averaged over
  $10^4$ initial conditions), in reservoir only (purple circles) and
  hybrid scheme (green squares), as a function of the reservoir size
  $D_R$. The solid curves display the average over all realizations,
  already presented in Fig.~\ref{fig:3}. Notice that, in the
  reservoir only scheme, a number of outliers are present for
  $D_R\lesssim 400$, these correspond to ``failed'' networks that make
  poor medium term predictions or even fail to reproduce the climate.
  Remarkably, such failures are not observed in the hybrid
  scheme. Inset: $\langle E_S\rangle$, i.e. the average (over network
  realizations and $10^4$ initial conditions for each realizations)
  ($\log_{10}$)error at the end of the open loop versus the reservoir
  size $D_R$ for the reservoir only (purple curve) and the hybrid
  (green curve) scheme, respectively. Symbols display the
  synchronization error in each network realization. Notice that there
  are no realizations with strong departure from the average as
  observed in main panel for the predictability time: this shows that
  the predictability performance is not always liked to the
  synchronization error (see text for a further discussion). Data refer to the case $c=10$ and $\Delta t=0.1$. For hyperparameters see Appendix~\ref{app:hyper}.
  \label{fig:S1}}
\end{figure}

In Fig.~\ref{fig:5}
we show the $D_R$-dependence of the average (over reservoir
realizations and initial conditions) predictability time, $T_p$,
defined as the first time such that the error between the network
predicted and reference trajectory reaches the threshold value
$\Delta^*=0.4  \langle ||\bm X||^2\rangle^{1/2}$. For $D_R\gtrsim 450$,
the predictability time saturates while for smaller reservoirs it can
be about threefold smaller and, in addition, with large fluctuations
mainly due to unsuccessful predictions, i.e. instances in which the
network is unable to proper modeling the dynamics (see
  Fig~\ref{fig:S1}). Remarkably, implementing the hybrid scheme even
with a poorly performing predictor such as the truncated model, the
forecasting ability of the network improves considerably (as also
shown in Fig.~\ref{fig:5}). In particular, with the hybrid scheme, saturation is reached earlier (for $D_R\gtrsim
300$) and, for smaller reservoirs, the predictability time of the
hybrid scheme is longer. Moreover, the hybrid scheme is less prone to
failures even for small $D_R$, hence fluctuations are smaller
(see Fig.~\ref{fig:S1}). Note that the chosen hybrid
design ensures that the improvement is only due to reservoir
capability of building a better effective model, reducing the average
synchronization ($\log_{10}$)error $\langle E_S\rangle$ (see the insets of
Fig.~\ref{fig:5} and \ref{fig:S1}, and the discussion in Appendix~\ref{app:hybrid}) and
thus the error on the initial condition of the slow variables. Indeed,
in the inset of Fig.~\ref{fig:5} we also show the
slope predicted on the basis of the slow perturbation growth rate,
$\lambda_s$ \cite{boffetta1998extension}.

The above observations boil down to the fact that the
difference between hybrid and reservoir only approach disappears at
increasing $D_R$ as the same plateau values for both synchronization
error and predictability time are reached. In other terms, if the
reservoir is large enough, adding the extra information from the
imperfect model does not improve the model produced by the
network. These conclusions can be cast in a positive
  message by saying that using a physically informed model allows for
reducing the size of the reservoir to achieve a reasonable
predictability and hence an effective model of the dynamics
with a smaller network, which is important when considering
multiscale systems of high dimensionality.

We remark that in Fig.~\ref{fig:5} the predictability time $T_p$ was
made nondimensional either using the growth rate of the slow dynamics
$\lambda_s$ (left scale of Fig.~\ref{fig:5}), or using
  the Lyapunov exponent $\lambda_f$ of the full system (right scale)
  which is dominated by the fast dynamics. For large networks the
predictability time is as large as $5$ (finite size) Lyapunov times,
which corresponds to about $70$ Lyapunov times with respect to the
full dynamics. Such a remarkably long predictability
with respect to the fastest time scale is typical of multiscale
systems, where the maximal Lyapunov exponent does not say much for the
predictability of the slow degrees of freedom
\cite{lorenz1996predictability,aurell1996growth,cencini2013finite}.

Figures~\ref{fig:2}a, \ref{fig:3}a and \ref{fig:5}(inset) (see also the inset of Fig.~\ref{fig:S1}), show that it is hard reaching synchronization error below
$10^{-2}$. Even when this happens it does not improve the
predictability, as the error quickly (even faster than the Lyapunov
time) raises to  values $O(10^{-2})$. Indeed, such an error threshold corresponds
to the crossover scale between the fast- and slow-controlled regime of
the perturbation growth (see Fig.~2 in
Ref.~\cite{boffetta1998extension}). In other terms, pushing the error
below this value requires the reservoir to (partially) reconstruct
also the fast component dynamics. 

\subsection{The role of the synchronization error\label{sec:error}}

In the previous section, we have used the average
  predictability time, $T_p$, as a performance metrics. If we
  interpret the synchronization error (at the end of the open loop) as
  the error on the initial conditions, since the system is chaotic, we
  could naively think that reducing such error always enhances the
  predictability. Consequently, one can expect the size of such error
  to be another good performance metrics.  In the following, we show
  that this is only partially true.

Obviously, to achieve long term predictability the
  smallness of the synchronization error is a necessary condition.
  Indeed the (log)error at the end of open loop cycle, $E_S$, puts an
  upper limit to the predictability time as
\begin{equation}
T_p\lesssim\frac{1}{\lambda_{s}}[\log(\Delta^*)-E_S]\,,
\label{eq:tpred}
\end{equation}
as confirmed by the solid line in the inset of Fig.~\ref{fig:5}.
However, it is not otherwise very informative about the overall
performance.  The reason is that the value
of $E_S,$ which can also be seen as the average error on one step
predictions, does not provide information on the structural stability
of the dynamics. Indeed, for a variety of hyperparameters values, we
have observed low $E_S$ resulting in failed predictions: in other
terms the model built by the network is not effective in forecasting
and in reproducing the climate. In these cases, the network was unable
to generate a good effective model, as shown in Fig.~\ref{fig:S1}: this
typically happens for relatively small $D_R$ in the reservoir only
implementation.

In a less extreme scenario, the error $E_S$ can be deceptively lower
than the scale at which the dynamics has been properly reconstructed. This
latter case is relevant to the multiscale setting since, as outlined
at the end of the previous section, fast variable reconstruction is necessary to
push the initial error below a certain threshold. In some cases, we
did observe the synchronization error falling below the typical value
$10^{-2}$ but immediately jumping back to it, implying unstable fast
scale reconstruction (for instance, see $c=3$, $\Delta t=0.01$ in
Fig.~\ref{fig:3}a).

As a consequence of the two above observations, $E_S$ is an unreliable
metric for hyperparameters landscape exploration as well.
We also remark that, even if fast scales were modeled with proper
architecture and training time, and $E_S$ could be pushed below the
crossover with an actual boost in performance, such improvements would
not dramatically increase the predictability time of the slow variables,
since they are suppressed by the
global (and greater, as dominated by the fast degrees of freedom)
Lyapunov exponent. This situation as discussed above is typical of
multiscale systems.

\section{Conclusions\label{sec:end}}
We have shown that reservoir computing is a promising machine learning
tool for building effective, data-driven models for multiscale chaotic
systems able to provide, in some cases, predictions as
  good as those that can be obtained with a perfect model with error
on the initial conditions. Moreover, the simplicity of the system
allowed us to gain insights into the inner work of the reservoir
computing approach that, at least for large scale separation, is
building an effective model akin to that obtained by asymptotic
multiscale techniques.  Finally, the reservoir computing approach can
be reinforced by blending it with an imperfect predictor, making it to
perform well also with smaller reservoirs. While we have obtained
these results with a relatively simple two-timescale model, given the
success of previous applications to spatially extended systems
\cite{ottPRL}, we think the method should work also with more complex
high dimensional multiscale systems.  In the latter, it may be
necessary to consider multi reservoir architectures
\cite{carmichael2019analysis} in parallel \cite{ottPRL}.  Moreover,
reservoir computing can be used to directly predict unobserved degrees
of freedom \cite{lu2017reservoir}. Using this scheme and the ideas
developed in this work it would be interesting to explore the
possibility to build novel subgrid schemes for turbulent flows
\cite{meneveau2000scale,sagaut2006large} (see also
Ref.~\cite{wikner2020combining} for a very recent attempt in this
direction based on reservoir computing with hybrid implementation),
preliminary tests could be performed in shell models for turbulence
for which physics only informed approaches have been only partially
successful \cite{biferale2017optimal}.

\begin{acknowledgments}
We thanks L. Biferale for early interest in the project and useful
discussions. AV and FB acknowledge MIUR-PRIN2017 ``Coarse-grained
description for non-equilibrium systems and transport phenomena''
(CO-NEST).  We acknowledge the CINECA award for the availability of
high performance computing resources and support from the GPU-AI
expert group in CINECA.
\end{acknowledgments}

\appendix
\section{Details on the implementation \label{app:implementation}}

\subsection{Intra-reservoir (R-to-R) connectivity matrix $\mathbb{W}_R$}
The intra-reservoir connectivity matrix, $\mathbb{W}_R$,  is
generated by drawing each entry from the same distribution. Each
element is the product of two random variables
$\mathbb{W}_{ij}=a*b\,$: $a$ being a real uniformly distributed random
number in $[-1,\;1]$ and $b$ taking values $1$ or $0$ with probability
$P_d=d/D_R$ and $1-P_d$, respectively. Consequently, each row has, on
average, $d$ non zero elements. Since $D_R\gg d$, the number of
non null entries per row is essentially distributed according to a
Poisson distribution. As a last step, the maximal eigenvalue (in absolute value),
$\rho_{\mathrm{max}}(\mathbb{W})$ of the resulting matrix $\mathbb{W}$
is computed and the matrix is rescaled element wise so that its new
spectral radius matches the target value $\rho$, i.e.:
$$ \mathbb{W}_R=\mathbb{W}\frac{\rho}{\rho_{\mathrm{max}}(\mathbb{W})} $$

\subsection{Input-to-reservoir (I-to-R)  connectivity matrix $\mathbb{W}_I$}
The input to reservoir matrix $\mathbb{W}_I$ is generated in such a
way that each reservoir node is connected to a single input. For this
purpose, for each row $j$, a single element $n_j$, uniformly chosen
between $1$ and the input dimension $D_I$, is different from
zero. This means that the reservoir node $j$ is only connected to the
$n_j^{th}$ input node. The connection strength is randomly chosen in
$[-\sigma,\sigma]$ with uniform distribution.

\subsection{Optimization of the output matrix $\mathbb{W}_O$}
The output matrix $\mathbb{W}_O$ is obtained via optimization. As
explained in Sec.~\ref{sec:RC}, $\mathbb{W}_O$ should be
chosen so that the output $\bm v(t)=\mathbb{W}_O\,\bm r^*(t)$ is, on
average, as similar as possible to the input signal $\bm
u(t)$. Incidentally, we remark that the use of $\bm r^*$ instead of
simply $\bm r$ is relevant to achieve accurate forecasting and is
heuristically motivated by the need to add some nonlinearity in the
network \cite{jaeger2001echo}.  The particular choice we adopted,
$r^*_i=r_i$ or $r_i^2$ for $i$ odd or even, respectively was suggested
in Refs.~\cite{ottLyapunov,ottAttractor} in view of the symmetries of the
Lorenz model.

As for the optimization of $\mathbb{W}_O$, we require that it should
minimize the cost function
\begin{equation}
  \label{cost}
  \mathcal{L}=\frac{1}{T_t}\sum_{0\leq  t\leq T_t}\| \mathbb{W}_O\,r^*(t)-u(t)\|^2+\alpha \,\mbox{tr}[ \mathbb{W}_O\, \mathbb{W}_O^T]\,,
\end{equation}
where $^T$ denotes the transpose and $T_t$ is the length of the
training input, whose choice is discussed below.  We point out that
the sum appearing in Eq.~(\ref{cost}) is a delicate quantity: we have
observed that moderate errors compromise the final performance. For
this reason, the Kahan summation has been employed in order to boost
numerical accuracy. The solution of the minimization of
Eq.~(\ref{cost}),
$$ \mathbb{W}^{opt}_O\mbox{ so that }\left.\frac{d\mathcal{L}}{d \mathbb{W}_O^T}\right|_{\mathbb{W}^{opt}_O} =0
$$
 is
$$ \mathbb{W}^{opt}_O=\langle u\otimes r^{*T}
 \rangle\;(\alpha\,\mathbb{I}_{D_R}+\langle r^{*}\otimes
 r^{*T}\rangle)^{-1} $$ where $\langle\cdot \rangle$ denotes the
 empirical average $\frac{1}{T_t}\sum_t$, $\otimes$ denotes the outer
 product and $\mathbb{I}_{D_R}$ the $D_R\times D_R$ identity
 matrix. The addend proportional to $\alpha$ in Eq.~(\ref{cost})  is the
 Tikhonov term, which is a $L^2$ regularization on
 $\mathbb{W}_O$. The Tikhonov regularization improves the numerical
 stability of the inversion, which could be critical if the ratio
 between the largest and the smallest eigenvalues, $\rho_{\mathrm{max}}$ and
 $\rho_{\mathrm{min}}$, is too large and the latter would behave as a (numerical) null
 eigenvalue \cite{Note2}, as it is the case for the
 dynamics we are studying.
 Here we have used $\alpha=10^{-8.5}$ which empirically was found to lead to
 $\log_{10}(\rho_{\mathrm{max}}/\rho_{\mathrm{min}})\approx 10$.

\subsection{Synchronization time and length of the training input trajectory}

All results presented in this article have been
  obtained using training trajectories of length $T_{t}=500$.  We
remark that using values $100\leq T_t\leq 1000$ one can hardly notice
qualitative differences.  At low training times, failures
can be very diverse, ranging from tilted attractors
to periodic orbits or spurious fixed points. The chosen values of
$T_t$ have been tested to be in the range that guarantees long term
reconstruction of the attractor with proper hyperparameters. As for
the the synchronizing length, we have chosen $T_s=4$. Such value is
about four times larger than the time actually needed to achieve best
possible synchronization indeed, as shown in the gray shaded areas of
Figs.~\ref{fig:2}a and \ref{fig:3}a, the error $E$ saturates to $E_S$
in about a time unit.

\subsection{Numerical details}

The whole code has been implemented in \textit{python3}, with linear
algebra performed via \textit{numpy}. Numerical integration of the
coupled Lorenz model were performed via a $4^{th}$ order Runge Kutta
scheme.

\subsection{Fixing the hyperparameters \label{app:hyper}}
The architecture of a generic network is described by a number of
parameters, often dubbed hyperparameters, e.g.: the number of layers,
activation functions etc. While a proper design is always crucial, in
the reservoir computing paradigm, this issue is especially critical
due to the absence of global optimization via backpropagation. The
reservoir-to-reservoir and input-to-reservoir connectivity matrices,
as discussed above, are quenched stochastic variables, whose
distribution depends on four hyperparameters:
$$
\mbox{Net}\sim P(\sigma,\rho,d,D_R)\,,
$$
namely, the strength of the I-to-R connection matrix  $\sigma$, the spectral radius $\rho$ of the R-to-R connection matrix, the degree $d$ of the R-to-R connection graph, and the reservoir size $D_R$.
Once the distribution is chosen, there are two separate issues.

The first is that, for a given choice of $(\sigma,\rho,d)$, the
network should be self-averaging if its size $D_R$ is large
enough. Indeed, we see from Fig.~\ref{fig:S1} that the variability
between realizations decreases with $D_R$, as expected.

The second issue is the choice of the triple $(\sigma,\rho,d)$. In
general, the existence of large and nearly flat (for any reasonable
performance metrics) region of suitable hyperparameters implies the
robustness of the method. As for the problem we have presented, such
region exists, even though, in the case $\Delta t=0.1$, $c=10$,
moderate fine tuning of the hyperparameters did improve the final
result, allowing to even (moderately) outperform the fully informed
twin model, as shown in Fig.~\ref{fig:2}a.

\begin{figure}[b!]
  \includegraphics[width=1\columnwidth]{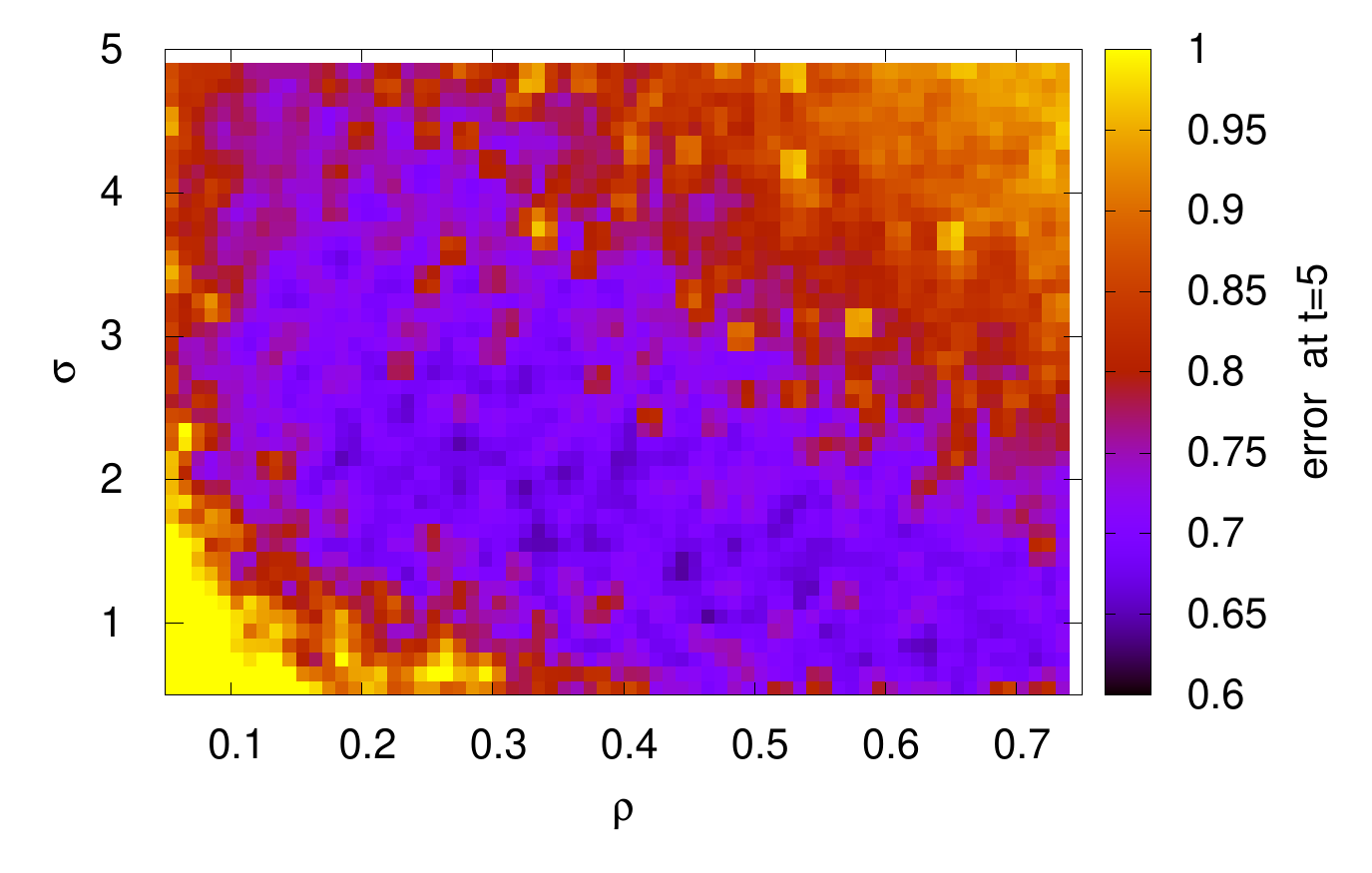}
  \caption{(Color online) Performance grid for $c=3$, $\Delta t=0.1$, $N=350$,
    $d=5$. Colors code error between forecasted and reference
    trajectory at time $t=5$ after closing the loop, which if the
    metrics here used $f=\|
    \boldsymbol{X}^{forecast}(t=5)-\boldsymbol{X}^{true}(t=5)\|$
    (averaged over 100 points of the attractor) for a single
    realization of the network for a given value of parameters
    $(\rho,\sigma)$. To highlight the suitable parameter region, a
    cutoff on has been put at $f=1$.}
  \label{fig:grid}
\end{figure}

It is important to remark that the characteristics of the regions of
suitable hyperparameters depend on the used metric. Here, we have
focused on medium term predictability, i.e. we evaluate the error
between forecasted and reference slow variables at a time (after
synchronization) that is much larger than one step $\Delta t$ but
before error saturation (corresponding to trajectories completely
uncorrelated).  Requiring a too short time predictability, as
discussed in Ref.~\cite{ottAttractor}, typically is not enough for
reproducing long time statistical properties of the target system
(i.e. the so called climate), as the learned attractor may be unstable
even if the dynamical equations are locally well approximated. If both
short term predictability and climate reproduction are required, the
suitable hyperparameter region typically shrinks. The metric we used
typically led to both predictability and climate reproduction, at
least for reservoir sizes large enough.

In order to fix the parameters, two techniques have been employed. The
first is the standard search on a grid (for a representative example
see Fig.~\ref{fig:grid}): a lattice is generated in the space of
parameters, then each node is evaluated according to some
cost function metrics. If such function is regular enough, it should
be possible to detect a suitable region of parameters. While this
default method is sound, it may require to train many independent
networks, even in poorly performing regions. Each network cannot be
too small for two reasons: the first is that small networks suffer
from higher inter realization fluctuations, the second is that we
cannot exclude that optimal $(\sigma,\rho,d)$ have a loose dependence
on the reservoir size $D_R$. As further discussed below we found a mild
dependence on the network degree $d$, provided it is not too large,
thus in Fig.\ref{fig:grid} we focused on the dependence on $\rho$ and
$\sigma$.

The second technique is the no gradient optimization method known as
particle swarming optimization (PSO)~\cite{kennedy1995particle}. PSO consists in
generating $n$ (we  used $n=10$) tuples of candidate -- the particles --
parameters, say ${\bf p}_i=(\rho_i,\sigma_i,d_i)
\;i=1,...,n$. At each step, each candidate is tested with a given
metrics $f$.  Here, we used the average (over $50-100$ initial
conditions) error on the slow variables after $t=2,4,5$ (depending on
the parameters) in the close loop configuration.  Then, at each
iteration $k$ of the algorithm, each candidate is accelerated towards
a stochastic mixture of its own best performing past position
$${\bf p}_i^*(k)=\arg\min_{{\bf p}_i(s)}\{f({\bf p}_i(s))|s<k\}$$
 and the overall best past performer 
 $${\bf p}^*(k)=\arg\min_{{{\bf p}_i^*}(k)}\{f({{\bf p}_i^*}(k))| i=1,...,n\}.$$ 
 Particles are evolved with the following second order time discrete dynamics  
 \begin{eqnarray}
 {\bf p}_i(k+1)&=&{\bf p}_i(k)+{\bf v}_i(k)
\nonumber \\
          {\bf v}_i(k+1)&=&w{\bf v}_i(k)+\phi^1_{i}(k)\,({\bf p}_i^*(k)-{\bf p}_i(k)) \nonumber\\
          &+&\phi^2_{i}(k)\,({\bf p}^*(k)-{\bf p}_i(k))
\nonumber
 \end{eqnarray}
with $\phi_i^j(k)\in[0,1]$ being random variables and $w\in[0,1]$
representing a form of inertia, as implemented in the python library
\textit{pyswarms}. After a suitable amount of iterations,
${\bf p}^*$ should be a valid candidate. The advantage of PSO is that,
after a transient, most candidate evaluations (each of which require
to initialize, train and test at least one network) should happen in
the good regions. It is worth pointing out that, unless self-averaging
is achieved thanks to large enough reservoir sizes, inter network
variability adds noise to limited attractor sampling when evaluating
$f$ and, therefore, fluctuations may appear and trap the algorithm in
suboptimal regions for some time. Moreover, the algorithm itself
depends on some hyperparameters that may have to be optimized
themselves by hand.
  
In our study, PSO has been mainly useful in fixing
parameters in the ($\Delta t=0.1$, $c=10$) case and to observe that
$d$ is the parameter which affects the performance the least. Some
gridding (especially in $\rho$ and $\sigma$) around the optimal
solution is useful, in general, as a cross check and to highlight the
robustness (or lack thereof) of the solution.

In Table~\ref{TablePar} we summarize the hyperparameters used in our
study.
\begin{table}[h!]
\begin{center}
\begin{tabular}{|c|c|c|}
\cline{2-3}
\multicolumn{1}{c|}{} & $\Delta t=0.1$ & $\Delta t=0.01$  \\
\hline
c=3& $\begin{matrix}d=5\\ \sigma=2\\ \rho=0.35\end{matrix}$ &  $\begin{matrix}d=5\\ \sigma=2.5\\ \rho=0.25\end{matrix}$   \\
\hline
c=10 &  $\begin{matrix}d=5\\ \sigma=1.8\\ \rho=0.34\end{matrix}$  &  $\begin{matrix}d=5\\ \sigma=0.8\\ \rho=0.68\end{matrix}$   \\
\hline
\end{tabular}
\end{center}
\caption{(Color online) Hyperparameters used in the simulations:
  $\Delta t$ is the sampling time, $c$ is the time scale separation of
  the multiscale scale Lorenz model
  Eqs.~(\ref{eq:lorenzslow}-\ref{eq:lorenzfast}), $\sigma$ is the
  input-to-reservoir coupling strength, $\rho$ is spectral radius of
  the reservoir-to-reservoir connectivity matrix and $d$ its degree. For the hybrid implementation, discussed in Sec.~\ref{sec:hybrid} and  Appendix~\ref{app:hybrid} we used the same hyperparameters.}
\label{TablePar}
\end{table}

\section{Multiscale model for the two time scales coupled Lorenz systems\label{app:model}}
In this Appendix we show how Eq.~(\ref{eq:improved}) was derived.
Following the notation of Eqs.~(\ref{eq:1}), we will denote with
$\boldsymbol{X}$ and $\boldsymbol{x}$ the slow $(X,Y,Z)$ and fast
$(x,z,y)$ variables, respectively. Our aim is to provide a model of
the fast variables in Eqs.~(\ref{eq:lorenzslow}) in terms of the slow
ones.  When the scale separation is very wide, we can
assume that $\boldsymbol{x}$ equilibrates, i.e. distribute according
to a stationary measure, for each value of the slow variables
$\boldsymbol{X}$ -- adiabatic principle --, and we call the expected
values with respect to such measure as
$\langle\cdot\rangle_{\boldsymbol{X}}$. We stress that
  the adiabatic approach requires a wide scale separation ($c\gg 1$)
  in order to work. In this limit, since only $Y$ enters the dynamics
of the fast variables, solely the value of $Y$ will
matter in building the adiabatic approximation,
i.e. $\langle\cdot\rangle_{\boldsymbol{X}}\equiv\langle\cdot\rangle_Y$. In
general, for moderate scale separation, this is not the case and a
``closure'' of the fast variables depending on the whole set of slow
variables would be required, a much harder task.

\begin{figure}[b!]
  \includegraphics[width=1\columnwidth]{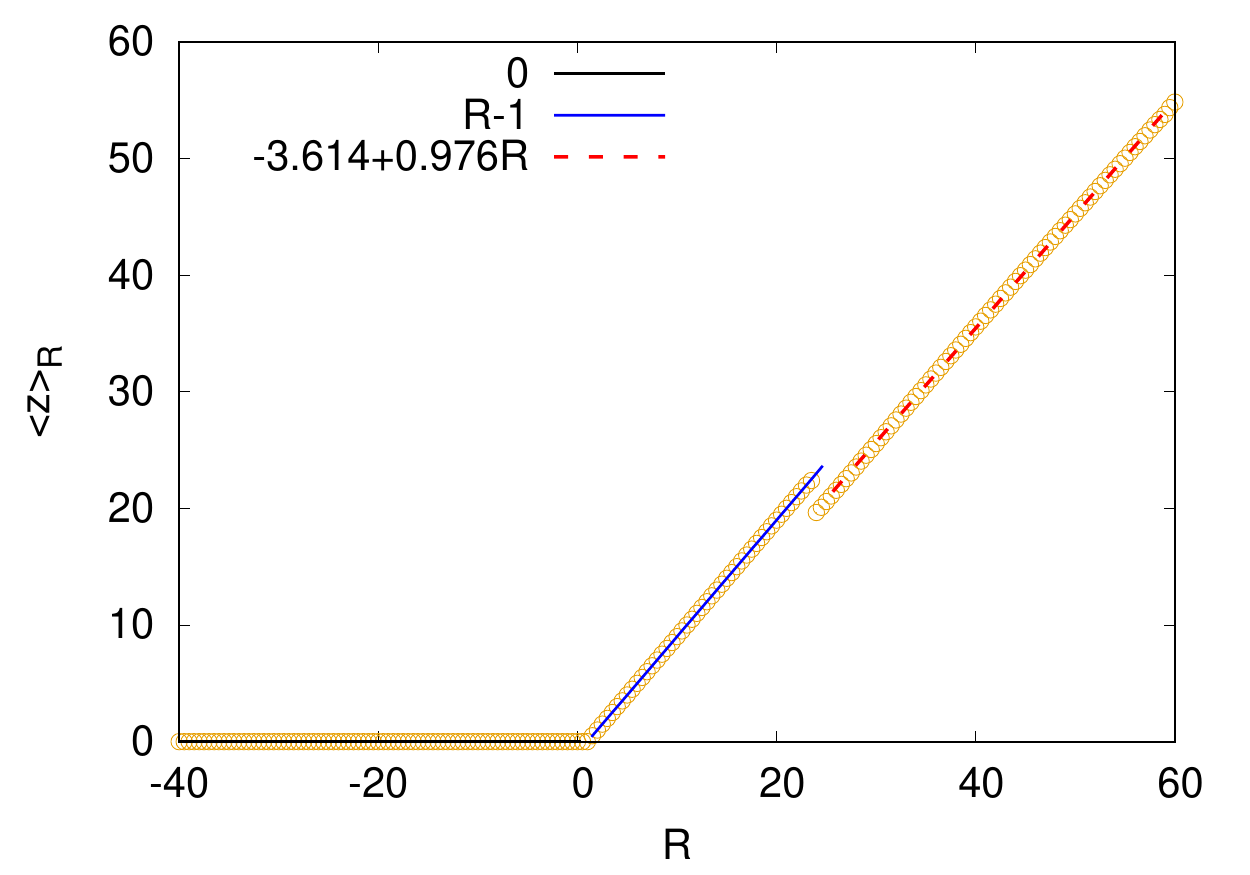}
  \caption{(Color online) $\langle z\rangle_R$ vs $R$ numerically computed in the standard Lorenz system (symbols). Notice that the curve is well approximated, in the range of interest, by the piecewise linear function in Eqs. (\ref{eq:piece}), see legend.}
  \label{fig:zvsR}
\end{figure}

\begin{figure}[t!]
  \includegraphics[width=1\columnwidth]{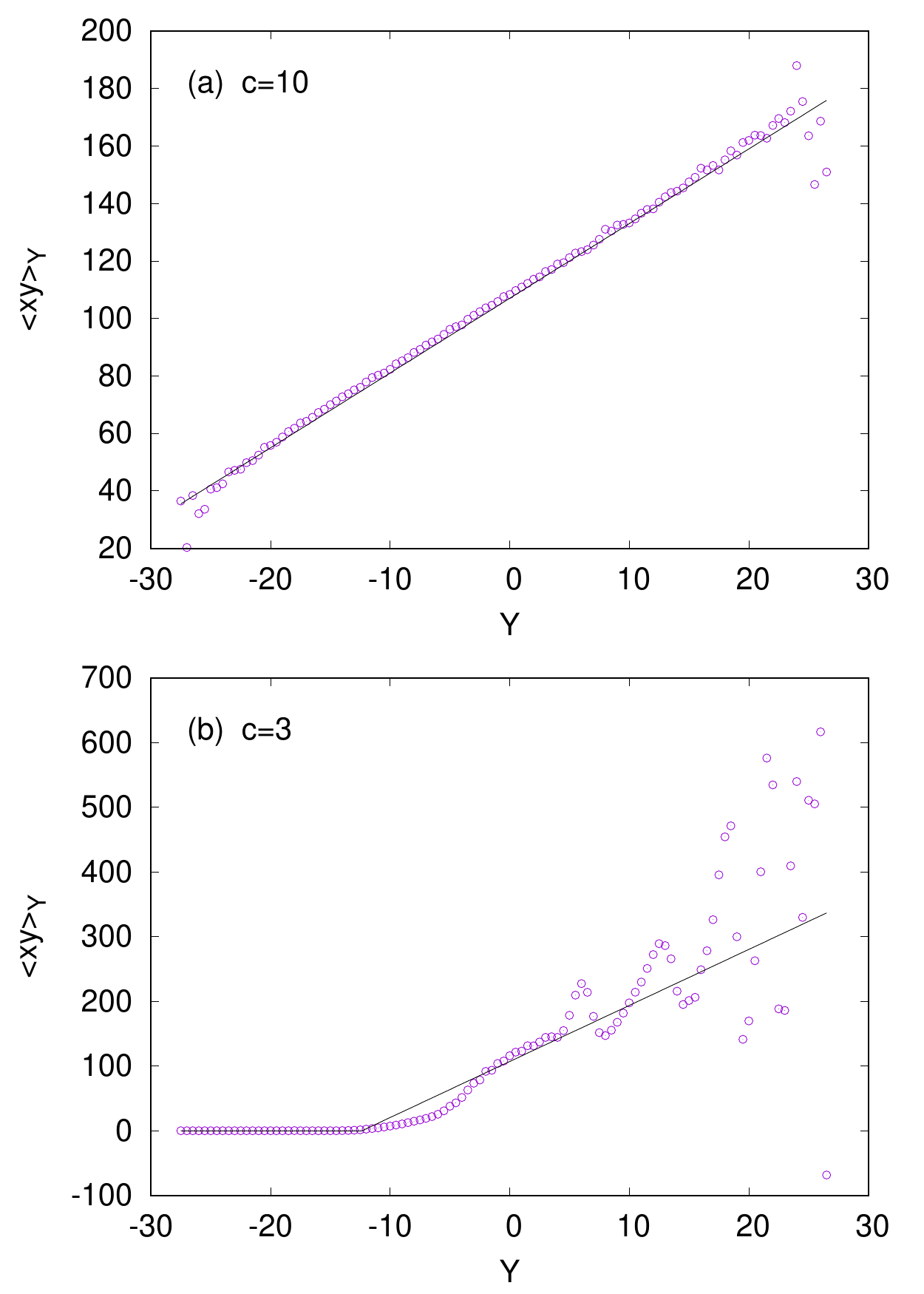}
  \caption{(Color online) Numerically computed conditional averages (symbols) $\langle xy|Y \rangle$
    for (a) $c=10$ and (b) $c=3$ and the corresponding multiscale (adiabatic)
    averages $\langle xy\rangle_Y$ given by Eqs.~(\ref{eq:improved2})
    (solid curve). For $c=10$ the two curves overlap. }
  \label{fig:ms}
\end{figure}

In order to model $\langle xy \rangle_{\boldsymbol{X}}$, we first
impose stationarity i.e. $\langle \dot{\bm
    x}\rangle_{\boldsymbol{X}}=0$ which, applied to the third
line of Eqs.~(\ref{eq:lorenzfast}), yields
\begin{equation}
  \langle xy\rangle_{\boldsymbol{X}}=b\langle z\rangle_{\boldsymbol{X}}\,.
  \label{eq:xy}
\end{equation}
Inserting the result (\ref{eq:xy}) in the equation for $Y$ in Eqs.~(\ref{eq:lorenzslow}) we obtain $\dot Y=R_sX-ZX-Y-\epsilon_s\,b\,\langle z\rangle_{\boldsymbol{X}}$.
Now we need to determine $\langle z\rangle_{\boldsymbol{X}}$. For this purpose, we notice that the equation for $\dot{y}$ in Eqs.~(\ref{eq:lorenzfast}) can be rewritten as
\begin{equation}
\dot{y}=c[Rx -zx-y]\qquad \mathrm{with}\qquad R=R_f+\frac{\epsilon_f}{c}Y \,.
\label{eq:newy}
\end{equation}
Exploiting the adiabatic principle we assume $Y$ (the slow variable) as fixed so that Eqs.~(\ref{eq:lorenzfast}) with the
second equation substituted with Eq.~(\ref{eq:newy}) become the
standard Lorenz model, a part from an inessential change of time
scale. Thus, we can now evolve the standard Lorenz
model and compute $\langle z \rangle_R$ (where $_R$ is just to remind
the $R$ dependence that will be reflected in a $Y$ dependence via
Eq.~(\ref{eq:newy})), which is shown in Fig.~\ref{fig:zvsR}.  As one can
see $\langle z \rangle_R$ depends on $R$ approximately as follows:
\begin{equation}
  \langle z \rangle_R \approx \left\{
\begin{array}{ll}
  0 & R<1\\
  R-1 & 1\leq R \lesssim 24.74\\
  0.976R-3.614 & R\gtrsim 24.74
\end{array}
\right.
\label{eq:piece}
\end{equation}
We remind that $R_c=24.74$ is the critical value at which the fixed point
\begin{equation}
z^* = (R-1)\Theta(R-1)\,,
\label{eq:fixed}
\end{equation}
(where $\Theta$ is the Heaviside step function)
loses its stability. Remarkably, $\langle z \rangle_R$ remain close to $z^*$ also for $R>R_c$.
The second expression in Eqs.~(\ref{eq:piece}), or equivalently Eq.~(\ref{eq:fixed}), yields
\begin{equation}
\langle xy \rangle_Y = b(R_f-1+(\epsilon_f/c)Y)\Theta(R_f-1+(\epsilon_f/c)Y)\,.
\label{ms_num}
\end{equation}
Using the numerical values of the constants ($b=8/3$, $R_f=45$ and
$\epsilon_f=10$), the above expression provides the estimate $\langle
xy \rangle_Y \approx (117.33+26.67\, Y/c) \,\Theta(117.33+26.67 Y/c)$  while using the third expression of
Eqs.~(\ref{eq:piece}) yields model (\ref{eq:improved}) that we used to compare with the network, i.e.
\begin{equation}
\langle xy\rangle_{Y}\!=\!(107.5\!+\!26.04Y/c) \,\Theta((107.5\!+\!26.04Y/c)\,.
  \label{eq:improved2}
\end{equation}
For $c=10$, the latter expression is very close to
the (numerically obtained) conditional average $\langle xy|Y \rangle
$(Fig.~\ref{fig:ms}a), confirming that the scale separation is
wide enough for the adiabatic approximation to work almost
perfectly. We notice that for $c=10$ the typical range of variation of $Y$ is such that $R$ mostly lies in the region of the third branch of Eqs.~(\ref{eq:piece}), explaining the validity of the approximation.

Conversely, for the case $c=3$, as shown in
Fig.~\ref{fig:ms}b the approximation is much cruder and important
deviations are present especially for large positive values of $Y$.
Indeed, in general,
 $$ \langle xy\rangle_{\boldsymbol{X}}\neq\langle
xy|\boldsymbol{X}\rangle$$ i.e. multiscale average obtained via the
adiabatic principle and the conditional average are not equivalent
since, in general,
$\langle\dot{\boldsymbol{x}}|\boldsymbol{X}\rangle\neq 0$. In this
case the values of all slow variables will matter in building a proper
effective model, a hard task even for the simple Lorenz model here
considered. However, as shown in Fig.~\ref{fig:3}, even in this case
the reservoir computing approach is quite performing even though it is
not straightforward to decipher the model it was able to devise.

\section{Discussion on various hybrid schemes implementations\label{app:hybrid}}

The hybrid scheme discussed in
  Sec.~\ref{sec:hybrid} allows for highlighting the properties of the
reservoir, but it is just one among the possible
  choices. Here, we briefly discuss three general schemes.

Let us assume, for simplicity, that our dynamical system, with state
variables $\bm s=(s_1,...,s_n)$, is described by the equation $\bm
s(t+1)=\bm f(\bm s(t))$, which is unknown. Here, without loss of
generality, we use discrete time dynamics and that we want to forecast
the whole set of state variables, this is just for the sake of
simplicity of the presentation.  Provided we have an imperfect model, $\bm
s(t+1)\approx \bm f_m(\bm s(t))$, for its evolution, we have basically
three options for building a hybrid scheme.

A first possibility is to approximate via machine learning only the
part of the signal that is not captured by the model $\bm f_m(\bm s(t))$. In other terms, one writes a forecast as
\begin{equation}
  \hat{\bm s}(t+1)=\bm f_m(\bm s(t))+\bm \delta_n(\bm s(t))
\end{equation}
where the residual $\bm \delta_n$ is given by the network, and can be
learned from a set of input-output pairs $\{\bm s(t),\bm s(t+1)-\bm
f_m(s(t))\}_{t=-T}^0$ according to some supervised learning
algorithm. In our framework, the hybrid network 
should be trained with the usual input but with target output given by
the difference between the true value of $\bm s(t+1)$ and the model
forecast $\bm f_m(\bm s(t))$.
\begin{figure}[b!]
  \includegraphics[width=1\columnwidth]{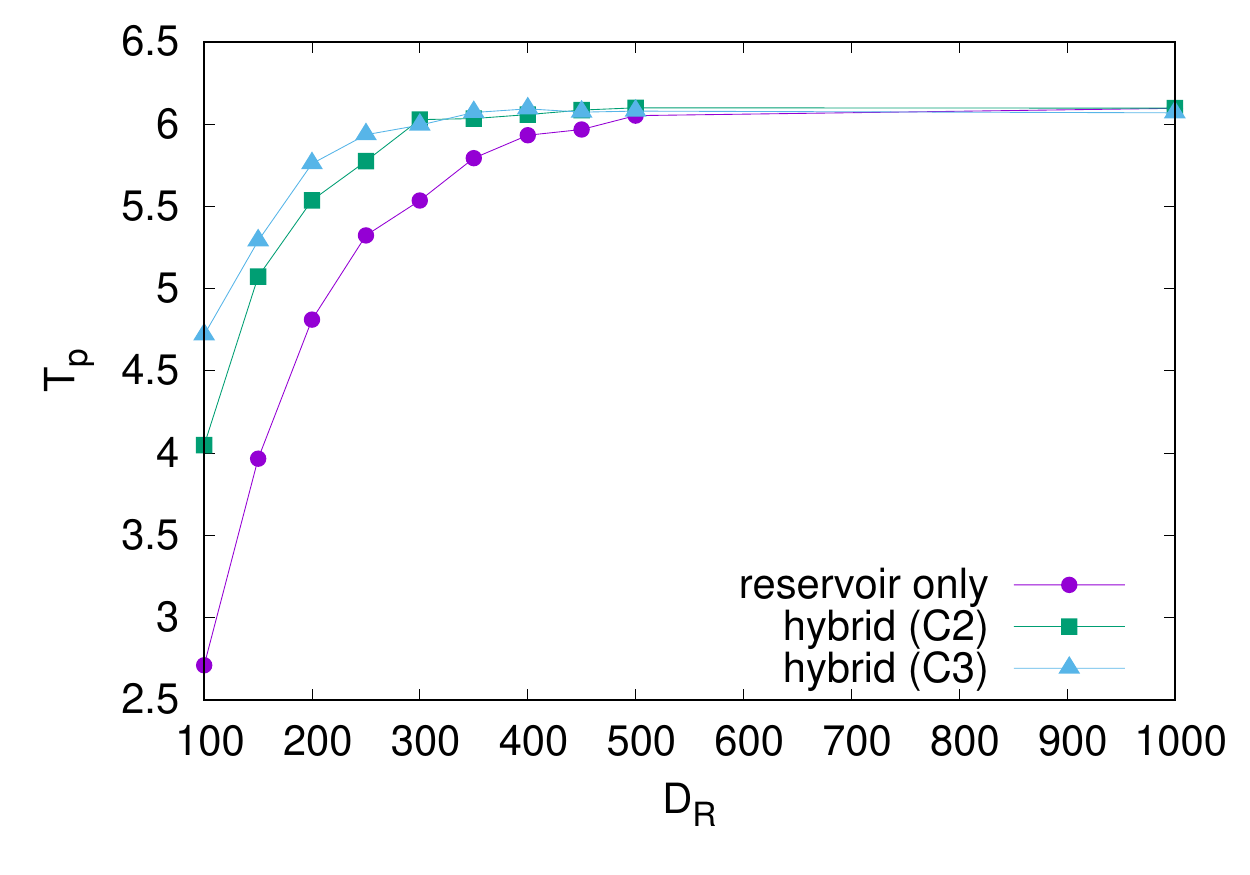}
  \caption{(Color online) Average (over $10^4$ initial conditions)
    predictability times are shown for reservoir only and two hybrid
    implementations ($\Delta t=0.1$ and $c=10$). The green line
    corresponds to the hybrid scheme (\ref{hyb1}), blue
    lines to the hybrid scheme (\ref{hyb2}) and purple lines to
    the reservoir only baseline. }
  \label{fig:hybrids}
\end{figure}

A second possibility is to add the available model prediction $\bm
f_m(\bm s(t))$ to the input $\bm s(t)$, obtaining an augmented input
$(\bm s(t),\bm f_m(\bm s(t))$ for the network. In this case, the
forecast reads as
\begin{equation}
   \label{hyb1}
   \hat{\bm s}(t+1)=\bm f_n(\bm s(t),\bm f_m(\bm s(t)).
\end{equation}
Clearly, if the model based prediction is very accurate, the network
will try to approximate the identity function. The network should be
trained with a set of input-outputs pairs
$\{(s(t),f_m(s(t))),s(t+1)\}_{t=-T}^0$. This is the approach we have
implemented in this article, in order to evaluate the performance of
the reservoir.

A third possibility is to  combine the two previous options, which is the approach followed in Ref.~\cite{ottHybrid}. In this case, the forecast  is obtained as:
\begin{equation}
  \label{hyb2}
  \hat {\bm s}(t+1)=\mathbb{A} \,\bm f_m(\bm s(t))+\mathbb{B}\,\bm \delta_n(\bm s(t),\bm f_m(s(t))\,,
\end{equation}
where the matrices $\mathbb{A}$ and $\mathbb{B}$ should be optimized,
along with $\delta_n$. This last option is a special case of the
second scheme, describing a residual multilayered neural network with
a linear output layer.

For the sake of completeness, in Fig.~\ref{fig:hybrids} we show how
this last architectures compares with the one we used in
Figs.~\ref{fig:5} and \ref{fig:S1} in terms of predictability. It
consists in taking the optimized linear combination of the predictions
from the hybrid net and the imperfect model. Namely, one augment the
$r^*$ array as $ \tilde {\bm r}^*=(\bm r^*, \bm f_m)$ and
then optimizes $\mathbb{W}_{O}$ to achieve $\bm
v(t+1)=\hat {\bm s}(t+1)\approx\mathbb{W}_{O}\tilde {\bm r}^*(t)$. As
one can see, the main effect is to slightly shift the
predictability-vs-size curve leftward, meaning that optimal
performance can be achieved with a slightly smaller network. However,
the improvement quickly disappears when the reservoir size increases.


\begin{thebibliography}{0}%
\makeatletter
\providecommand \@ifxundefined [1]{%
 \@ifx{#1\undefined}
}%
\providecommand \@ifnum [1]{%
 \ifnum #1\expandafter \@firstoftwo
 \else \expandafter \@secondoftwo
 \fi
}%
\providecommand \@ifx [1]{%
 \ifx #1\expandafter \@firstoftwo
 \else \expandafter \@secondoftwo
 \fi
}%
\providecommand \natexlab [1]{#1}%
\providecommand \enquote  [1]{``#1''}%
\providecommand \bibnamefont  [1]{#1}%
\providecommand \bibfnamefont [1]{#1}%
\providecommand \citenamefont [1]{#1}%
\providecommand \href@noop [0]{\@secondoftwo}%
\providecommand \href [0]{\begingroup \@sanitize@url \@href}%
\providecommand \@href[1]{\@@startlink{#1}\@@href}%
\providecommand \@@href[1]{\endgroup#1\@@endlink}%
\providecommand \@sanitize@url [0]{\catcode `\\12\catcode `\$12\catcode
  `\&12\catcode `\#12\catcode `\^12\catcode `\_12\catcode `\%12\relax}%
\providecommand \@@startlink[1]{}%
\providecommand \@@endlink[0]{}%
\providecommand \url  [0]{\begingroup\@sanitize@url \@url }%
\providecommand \@url [1]{\endgroup\@href {#1}{\urlprefix }}%
\providecommand \urlprefix  [0]{URL }%
\providecommand \Eprint [0]{\href }%
\providecommand \doibase [0]{https://doi.org/}%
\providecommand \selectlanguage [0]{\@gobble}%
\providecommand \bibinfo  [0]{\@secondoftwo}%
\providecommand \bibfield  [0]{\@secondoftwo}%
\providecommand \translation [1]{[#1]}%
\providecommand \BibitemOpen [0]{}%
\providecommand \bibitemStop [0]{}%
\providecommand \bibitemNoStop [0]{.\EOS\space}%
\providecommand \EOS [0]{\spacefactor3000\relax}%
\providecommand \BibitemShut  [1]{\csname bibitem#1\endcsname}%
\let\auto@bib@innerbib\@empty
\end{thebibliography}%


\begin{thebibliography}{44}%
\makeatletter
\providecommand \@ifxundefined [1]{%
 \@ifx{#1\undefined}
}%
\providecommand \@ifnum [1]{%
 \ifnum #1\expandafter \@firstoftwo
 \else \expandafter \@secondoftwo
 \fi
}%
\providecommand \@ifx [1]{%
 \ifx #1\expandafter \@firstoftwo
 \else \expandafter \@secondoftwo
 \fi
}%
\providecommand \natexlab [1]{#1}%
\providecommand \enquote  [1]{``#1''}%
\providecommand \bibnamefont  [1]{#1}%
\providecommand \bibfnamefont [1]{#1}%
\providecommand \citenamefont [1]{#1}%
\providecommand \href@noop [0]{\@secondoftwo}%
\providecommand \href [0]{\begingroup \@sanitize@url \@href}%
\providecommand \@href[1]{\@@startlink{#1}\@@href}%
\providecommand \@@href[1]{\endgroup#1\@@endlink}%
\providecommand \@sanitize@url [0]{\catcode `\\12\catcode `\$12\catcode
  `\&12\catcode `\#12\catcode `\^12\catcode `\_12\catcode `\%12\relax}%
\providecommand \@@startlink[1]{}%
\providecommand \@@endlink[0]{}%
\providecommand \url  [0]{\begingroup\@sanitize@url \@url }%
\providecommand \@url [1]{\endgroup\@href {#1}{\urlprefix }}%
\providecommand \urlprefix  [0]{URL }%
\providecommand \Eprint [0]{\href }%
\providecommand \doibase [0]{http://dx.doi.org/}%
\providecommand \selectlanguage [0]{\@gobble}%
\providecommand \bibinfo  [0]{\@secondoftwo}%
\providecommand \bibfield  [0]{\@secondoftwo}%
\providecommand \translation [1]{[#1]}%
\providecommand \BibitemOpen [0]{}%
\providecommand \bibitemStop [0]{}%
\providecommand \bibitemNoStop [0]{.\EOS\space}%
\providecommand \EOS [0]{\spacefactor3000\relax}%
\providecommand \BibitemShut  [1]{\csname bibitem#1\endcsname}%
\let\auto@bib@innerbib\@empty
\bibitem [{\citenamefont {Argall}\ \emph {et~al.}(2009)\citenamefont {Argall},
  \citenamefont {Chernova}, \citenamefont {Veloso},\ and\ \citenamefont
  {Browning}}]{argall2009survey}%
  \BibitemOpen
  \bibfield  {author} {\bibinfo {author} {\bibfnamefont {B.~D.}\ \bibnamefont
  {Argall}}, \bibinfo {author} {\bibfnamefont {S.}~\bibnamefont {Chernova}},
  \bibinfo {author} {\bibfnamefont {M.}~\bibnamefont {Veloso}}, \ and\ \bibinfo
  {author} {\bibfnamefont {B.}~\bibnamefont {Browning}},\ }\href@noop {}
  {\bibfield  {journal} {\bibinfo  {journal} {Robot. Auton. Sys.}\ }\textbf
  {\bibinfo {volume} {57}},\ \bibinfo {pages} {469} (\bibinfo {year}
  {2009})}\BibitemShut {NoStop}%
\bibitem [{\citenamefont {Libbrecht}\ and\ \citenamefont
  {Noble}(2015)}]{libbrecht2015machine}%
  \BibitemOpen
  \bibfield  {author} {\bibinfo {author} {\bibfnamefont {M.~W.}\ \bibnamefont
  {Libbrecht}}\ and\ \bibinfo {author} {\bibfnamefont {W.~S.}\ \bibnamefont
  {Noble}},\ }\href@noop {} {\bibfield  {journal} {\bibinfo  {journal} {Nature
  Rev. Genet.}\ }\textbf {\bibinfo {volume} {16}},\ \bibinfo {pages} {321}
  (\bibinfo {year} {2015})}\BibitemShut {NoStop}%
\bibitem [{\citenamefont {He}\ \emph {et~al.}(2019)\citenamefont {He},
  \citenamefont {Baxter}, \citenamefont {Xu}, \citenamefont {Xu}, \citenamefont
  {Zhou},\ and\ \citenamefont {Zhang}}]{he2019practical}%
  \BibitemOpen
  \bibfield  {author} {\bibinfo {author} {\bibfnamefont {J.}~\bibnamefont
  {He}}, \bibinfo {author} {\bibfnamefont {S.~L.}\ \bibnamefont {Baxter}},
  \bibinfo {author} {\bibfnamefont {J.}~\bibnamefont {Xu}}, \bibinfo {author}
  {\bibfnamefont {J.}~\bibnamefont {Xu}}, \bibinfo {author} {\bibfnamefont
  {X.}~\bibnamefont {Zhou}}, \ and\ \bibinfo {author} {\bibfnamefont
  {K.}~\bibnamefont {Zhang}},\ }\href@noop {} {\bibfield  {journal} {\bibinfo
  {journal} {Nature Med.}\ }\textbf {\bibinfo {volume} {25}},\ \bibinfo {pages}
  {30} (\bibinfo {year} {2019})}\BibitemShut {NoStop}%
\bibitem [{\citenamefont {Carleo}\ \emph {et~al.}(2019)\citenamefont {Carleo},
  \citenamefont {Cirac}, \citenamefont {Cranmer}, \citenamefont {Daudet},
  \citenamefont {Schuld}, \citenamefont {Tishby}, \citenamefont
  {Vogt-Maranto},\ and\ \citenamefont {Zdeborov{\'a}}}]{carleo2019machine}%
  \BibitemOpen
  \bibfield  {author} {\bibinfo {author} {\bibfnamefont {G.}~\bibnamefont
  {Carleo}}, \bibinfo {author} {\bibfnamefont {I.}~\bibnamefont {Cirac}},
  \bibinfo {author} {\bibfnamefont {K.}~\bibnamefont {Cranmer}}, \bibinfo
  {author} {\bibfnamefont {L.}~\bibnamefont {Daudet}}, \bibinfo {author}
  {\bibfnamefont {M.}~\bibnamefont {Schuld}}, \bibinfo {author} {\bibfnamefont
  {N.}~\bibnamefont {Tishby}}, \bibinfo {author} {\bibfnamefont
  {L.}~\bibnamefont {Vogt-Maranto}}, \ and\ \bibinfo {author} {\bibfnamefont
  {L.}~\bibnamefont {Zdeborov{\'a}}},\ }\href@noop {} {\bibfield  {journal}
  {\bibinfo  {journal} {Rev. Mod. Phys.}\ }\textbf {\bibinfo {volume} {91}},\
  \bibinfo {pages} {045002} (\bibinfo {year} {2019})}\BibitemShut {NoStop}%
\bibitem [{\citenamefont {Verstraeten}\ \emph {et~al.}(2007)\citenamefont
  {Verstraeten}, \citenamefont {Schrauwen}, \citenamefont {d’Haene},\ and\
  \citenamefont {Stroobandt}}]{verstraeten2007experimental}%
  \BibitemOpen
  \bibfield  {author} {\bibinfo {author} {\bibfnamefont {D.}~\bibnamefont
  {Verstraeten}}, \bibinfo {author} {\bibfnamefont {B.}~\bibnamefont
  {Schrauwen}}, \bibinfo {author} {\bibfnamefont {M.}~\bibnamefont
  {d’Haene}}, \ and\ \bibinfo {author} {\bibfnamefont {D.}~\bibnamefont
  {Stroobandt}},\ }\href@noop {} {\bibfield  {journal} {\bibinfo  {journal}
  {Neural Net.}\ }\textbf {\bibinfo {volume} {20}},\ \bibinfo {pages} {391}
  (\bibinfo {year} {2007})}\BibitemShut {NoStop}%
\bibitem [{\citenamefont {Schrauwen}\ \emph {et~al.}(2007)\citenamefont
  {Schrauwen}, \citenamefont {Verstraeten},\ and\ \citenamefont
  {Van~Campenhout}}]{schrauwen2007overview}%
  \BibitemOpen
  \bibfield  {author} {\bibinfo {author} {\bibfnamefont {B.}~\bibnamefont
  {Schrauwen}}, \bibinfo {author} {\bibfnamefont {D.}~\bibnamefont
  {Verstraeten}}, \ and\ \bibinfo {author} {\bibfnamefont {J.}~\bibnamefont
  {Van~Campenhout}},\ }in\ \href@noop {} {\emph {\bibinfo {booktitle} {Proc.
  15th Europ. Symp. Artif. Neural Netw.}}}\ (\bibinfo {year} {2007})\ pp.\
  \bibinfo {pages} {471--482}\BibitemShut {NoStop}%
\bibitem [{\citenamefont {Jaeger}(2001)}]{jaeger2001echo}%
  \BibitemOpen
  \bibfield  {author} {\bibinfo {author} {\bibfnamefont {H.}~\bibnamefont
  {Jaeger}},\ }\href@noop {} {\bibfield  {journal} {\bibinfo  {journal} {German
  Nat. Res. Center Infor. Tech. GMD Technical Report}\ }\textbf {\bibinfo
  {volume} {148}},\ \bibinfo {pages} {13} (\bibinfo {year} {2001})}\BibitemShut
  {NoStop}%
\bibitem [{\citenamefont {Luko{\v{s}}evi{\v{c}}ius}\ and\ \citenamefont
  {Jaeger}(2009)}]{JaegerReview}%
  \BibitemOpen
  \bibfield  {author} {\bibinfo {author} {\bibfnamefont {M.}~\bibnamefont
  {Luko{\v{s}}evi{\v{c}}ius}}\ and\ \bibinfo {author} {\bibfnamefont
  {H.}~\bibnamefont {Jaeger}},\ }\href@noop {} {\bibfield  {journal} {\bibinfo
  {journal} {Comp. Sci. Rev.}\ }\textbf {\bibinfo {volume} {3}},\ \bibinfo
  {pages} {127} (\bibinfo {year} {2009})}\BibitemShut {NoStop}%
\bibitem [{\citenamefont {Jaeger}\ and\ \citenamefont
  {Haas}(2004)}]{jaegerScience}%
  \BibitemOpen
  \bibfield  {author} {\bibinfo {author} {\bibfnamefont {H.}~\bibnamefont
  {Jaeger}}\ and\ \bibinfo {author} {\bibfnamefont {H.}~\bibnamefont {Haas}},\
  }\href@noop {} {\bibfield  {journal} {\bibinfo  {journal} {Science}\ }\textbf
  {\bibinfo {volume} {304}},\ \bibinfo {pages} {78} (\bibinfo {year}
  {2004})}\BibitemShut {NoStop}%
\bibitem [{\citenamefont {Pathak}\ \emph {et~al.}(2017)\citenamefont {Pathak},
  \citenamefont {Lu}, \citenamefont {Hunt}, \citenamefont {Girvan},\ and\
  \citenamefont {Ott}}]{ottLyapunov}%
  \BibitemOpen
  \bibfield  {author} {\bibinfo {author} {\bibfnamefont {J.}~\bibnamefont
  {Pathak}}, \bibinfo {author} {\bibfnamefont {Z.}~\bibnamefont {Lu}}, \bibinfo
  {author} {\bibfnamefont {B.~R.}\ \bibnamefont {Hunt}}, \bibinfo {author}
  {\bibfnamefont {M.}~\bibnamefont {Girvan}}, \ and\ \bibinfo {author}
  {\bibfnamefont {E.}~\bibnamefont {Ott}},\ }\href@noop {} {\bibfield
  {journal} {\bibinfo  {journal} {Chaos}\ }\textbf {\bibinfo {volume} {27}},\
  \bibinfo {pages} {121102} (\bibinfo {year} {2017})}\BibitemShut {NoStop}%
\bibitem [{\citenamefont {Pathak}\ \emph
  {et~al.}(2018{\natexlab{a}})\citenamefont {Pathak}, \citenamefont {Hunt},
  \citenamefont {Girvan}, \citenamefont {Lu},\ and\ \citenamefont
  {Ott}}]{ottPRL}%
  \BibitemOpen
  \bibfield  {author} {\bibinfo {author} {\bibfnamefont {J.}~\bibnamefont
  {Pathak}}, \bibinfo {author} {\bibfnamefont {B.}~\bibnamefont {Hunt}},
  \bibinfo {author} {\bibfnamefont {M.}~\bibnamefont {Girvan}}, \bibinfo
  {author} {\bibfnamefont {Z.}~\bibnamefont {Lu}}, \ and\ \bibinfo {author}
  {\bibfnamefont {E.}~\bibnamefont {Ott}},\ }\href@noop {} {\bibfield
  {journal} {\bibinfo  {journal} {Phys. Rev. Lett.}\ }\textbf {\bibinfo
  {volume} {120}},\ \bibinfo {pages} {024102} (\bibinfo {year}
  {2018}{\natexlab{a}})}\BibitemShut {NoStop}%
\bibitem [{\citenamefont {Lu}\ \emph {et~al.}(2018)\citenamefont {Lu},
  \citenamefont {Hunt},\ and\ \citenamefont {Ott}}]{ottAttractor}%
  \BibitemOpen
  \bibfield  {author} {\bibinfo {author} {\bibfnamefont {Z.}~\bibnamefont
  {Lu}}, \bibinfo {author} {\bibfnamefont {B.~R.}\ \bibnamefont {Hunt}}, \ and\
  \bibinfo {author} {\bibfnamefont {E.}~\bibnamefont {Ott}},\ }\href@noop {}
  {\bibfield  {journal} {\bibinfo  {journal} {Chaos}\ }\textbf {\bibinfo
  {volume} {28}},\ \bibinfo {pages} {061104} (\bibinfo {year}
  {2018})}\BibitemShut {NoStop}%
\bibitem [{\citenamefont {Vlachas}\ \emph {et~al.}(2018)\citenamefont
  {Vlachas}, \citenamefont {Byeon}, \citenamefont {Wan}, \citenamefont
  {Sapsis},\ and\ \citenamefont {Koumoutsakos}}]{vlachas2018data}%
  \BibitemOpen
  \bibfield  {author} {\bibinfo {author} {\bibfnamefont {P.~R.}\ \bibnamefont
  {Vlachas}}, \bibinfo {author} {\bibfnamefont {W.}~\bibnamefont {Byeon}},
  \bibinfo {author} {\bibfnamefont {Z.~Y.}\ \bibnamefont {Wan}}, \bibinfo
  {author} {\bibfnamefont {T.~P.}\ \bibnamefont {Sapsis}}, \ and\ \bibinfo
  {author} {\bibfnamefont {P.}~\bibnamefont {Koumoutsakos}},\ }\href@noop {}
  {\bibfield  {journal} {\bibinfo  {journal} {Proc. Royal Soc. A}\ }\textbf
  {\bibinfo {volume} {474}},\ \bibinfo {pages} {20170844} (\bibinfo {year}
  {2018})}\BibitemShut {NoStop}%
\bibitem [{\citenamefont {Nakai}\ and\ \citenamefont
  {Saiki}(2018)}]{nakai2018machine}%
  \BibitemOpen
  \bibfield  {author} {\bibinfo {author} {\bibfnamefont {K.}~\bibnamefont
  {Nakai}}\ and\ \bibinfo {author} {\bibfnamefont {Y.}~\bibnamefont {Saiki}},\
  }\href@noop {} {\bibfield  {journal} {\bibinfo  {journal} {Phys. Rev. E}\
  }\textbf {\bibinfo {volume} {98}},\ \bibinfo {pages} {023111} (\bibinfo
  {year} {2018})}\BibitemShut {NoStop}%
\bibitem [{\citenamefont {Pathak}\ \emph
  {et~al.}(2018{\natexlab{b}})\citenamefont {Pathak}, \citenamefont {Wikner},
  \citenamefont {Fussell}, \citenamefont {Chandra}, \citenamefont {Hunt},
  \citenamefont {Girvan},\ and\ \citenamefont {Ott}}]{ottHybrid}%
  \BibitemOpen
  \bibfield  {author} {\bibinfo {author} {\bibfnamefont {J.}~\bibnamefont
  {Pathak}}, \bibinfo {author} {\bibfnamefont {A.}~\bibnamefont {Wikner}},
  \bibinfo {author} {\bibfnamefont {R.}~\bibnamefont {Fussell}}, \bibinfo
  {author} {\bibfnamefont {S.}~\bibnamefont {Chandra}}, \bibinfo {author}
  {\bibfnamefont {B.~R.}\ \bibnamefont {Hunt}}, \bibinfo {author}
  {\bibfnamefont {M.}~\bibnamefont {Girvan}}, \ and\ \bibinfo {author}
  {\bibfnamefont {E.}~\bibnamefont {Ott}},\ }\href@noop {} {\bibfield
  {journal} {\bibinfo  {journal} {Chaos}\ }\textbf {\bibinfo {volume} {28}},\
  \bibinfo {pages} {041101} (\bibinfo {year} {2018}{\natexlab{b}})}\BibitemShut
  {NoStop}%
\bibitem [{\citenamefont {Wikner}\ \emph {et~al.}(2020)\citenamefont {Wikner},
  \citenamefont {Pathak}, \citenamefont {Hunt}, \citenamefont {Girvan},
  \citenamefont {Arcomano}, \citenamefont {Szunyogh}, \citenamefont
  {Pomerance},\ and\ \citenamefont {Ott}}]{wikner2020combining}%
  \BibitemOpen
  \bibfield  {author} {\bibinfo {author} {\bibfnamefont {A.}~\bibnamefont
  {Wikner}}, \bibinfo {author} {\bibfnamefont {J.}~\bibnamefont {Pathak}},
  \bibinfo {author} {\bibfnamefont {B.}~\bibnamefont {Hunt}}, \bibinfo {author}
  {\bibfnamefont {M.}~\bibnamefont {Girvan}}, \bibinfo {author} {\bibfnamefont
  {T.}~\bibnamefont {Arcomano}}, \bibinfo {author} {\bibfnamefont
  {I.}~\bibnamefont {Szunyogh}}, \bibinfo {author} {\bibfnamefont
  {A.}~\bibnamefont {Pomerance}}, \ and\ \bibinfo {author} {\bibfnamefont
  {E.}~\bibnamefont {Ott}},\ }\href@noop {} {\bibfield  {journal} {\bibinfo
  {journal} {Chaos}\ }\textbf {\bibinfo {volume} {30}},\ \bibinfo {pages}
  {053111} (\bibinfo {year} {2020})}\BibitemShut {NoStop}%
\bibitem [{\citenamefont {Warhaft}(2002)}]{warhaft2002turbulence}%
  \BibitemOpen
  \bibfield  {author} {\bibinfo {author} {\bibfnamefont {Z.}~\bibnamefont
  {Warhaft}},\ }\href@noop {} {\bibfield  {journal} {\bibinfo  {journal} {Proc.
  Nat. Acad. Sci.}\ }\textbf {\bibinfo {volume} {99}},\ \bibinfo {pages} {2481}
  (\bibinfo {year} {2002})}\BibitemShut {NoStop}%
\bibitem [{\citenamefont {Peixoto}\ and\ \citenamefont
  {Oort}(1992)}]{peixoto1992physics}%
  \BibitemOpen
  \bibfield  {author} {\bibinfo {author} {\bibfnamefont {J.~P.}\ \bibnamefont
  {Peixoto}}\ and\ \bibinfo {author} {\bibfnamefont {A.~H.}\ \bibnamefont
  {Oort}},\ }\href@noop {} {\emph {\bibinfo {title} {Physics of climate}}}\
  (\bibinfo  {publisher} {New York, NY (United States); American Institute of
  Physics},\ \bibinfo {year} {1992})\BibitemShut {NoStop}%
\bibitem [{\citenamefont {Pedlosky}(2013)}]{pedlosky2013geophysical}%
  \BibitemOpen
  \bibfield  {author} {\bibinfo {author} {\bibfnamefont {J.}~\bibnamefont
  {Pedlosky}},\ }\href@noop {} {\emph {\bibinfo {title} {Geophysical fluid
  dynamics}}}\ (\bibinfo  {publisher} {Springer},\ \bibinfo {year}
  {2013})\BibitemShut {NoStop}%
\bibitem [{\citenamefont {Lorenz}(1995)}]{lorenz1996predictability}%
  \BibitemOpen
  \bibfield  {author} {\bibinfo {author} {\bibfnamefont {E.~N.}\ \bibnamefont
  {Lorenz}},\ }in\ \href@noop {} {\emph {\bibinfo {booktitle} {ECMWF Seminar
  Proceedings on Predictability}}},\ Vol.~\bibinfo {volume} {1}\ (\bibinfo
  {publisher} {ECMWF, Reading, UK},\ \bibinfo {year} {1995})\BibitemShut
  {NoStop}%
\bibitem [{\citenamefont {Aurell}\ \emph {et~al.}(1996)\citenamefont {Aurell},
  \citenamefont {Boffetta}, \citenamefont {Crisanti}, \citenamefont {Paladin},\
  and\ \citenamefont {Vulpiani}}]{aurell1996growth}%
  \BibitemOpen
  \bibfield  {author} {\bibinfo {author} {\bibfnamefont {E.}~\bibnamefont
  {Aurell}}, \bibinfo {author} {\bibfnamefont {G.}~\bibnamefont {Boffetta}},
  \bibinfo {author} {\bibfnamefont {A.}~\bibnamefont {Crisanti}}, \bibinfo
  {author} {\bibfnamefont {G.}~\bibnamefont {Paladin}}, \ and\ \bibinfo
  {author} {\bibfnamefont {A.}~\bibnamefont {Vulpiani}},\ }\href@noop {}
  {\bibfield  {journal} {\bibinfo  {journal} {Phys. Rev. Lett.}\ }\textbf
  {\bibinfo {volume} {77}},\ \bibinfo {pages} {1262} (\bibinfo {year}
  {1996})}\BibitemShut {NoStop}%
\bibitem [{\citenamefont {Cencini}\ and\ \citenamefont
  {Vulpiani}(2013)}]{cencini2013finite}%
  \BibitemOpen
  \bibfield  {author} {\bibinfo {author} {\bibfnamefont {M.}~\bibnamefont
  {Cencini}}\ and\ \bibinfo {author} {\bibfnamefont {A.}~\bibnamefont
  {Vulpiani}},\ }\href@noop {} {\bibfield  {journal} {\bibinfo  {journal} {J.
  Phys. A}\ }\textbf {\bibinfo {volume} {46}},\ \bibinfo {pages} {254019}
  (\bibinfo {year} {2013})}\BibitemShut {NoStop}%
\bibitem [{\citenamefont {Boffetta}\ \emph {et~al.}(2000)\citenamefont
  {Boffetta}, \citenamefont {Celani}, \citenamefont {Cencini}, \citenamefont
  {Lacorata},\ and\ \citenamefont {Vulpiani}}]{boffetta2000predictability}%
  \BibitemOpen
  \bibfield  {author} {\bibinfo {author} {\bibfnamefont {G.}~\bibnamefont
  {Boffetta}}, \bibinfo {author} {\bibfnamefont {A.}~\bibnamefont {Celani}},
  \bibinfo {author} {\bibfnamefont {M.}~\bibnamefont {Cencini}}, \bibinfo
  {author} {\bibfnamefont {G.}~\bibnamefont {Lacorata}}, \ and\ \bibinfo
  {author} {\bibfnamefont {A.}~\bibnamefont {Vulpiani}},\ }\href@noop {}
  {\bibfield  {journal} {\bibinfo  {journal} {J. Phys. A}\ }\textbf {\bibinfo
  {volume} {33}},\ \bibinfo {pages} {1313} (\bibinfo {year}
  {2000})}\BibitemShut {NoStop}%
\bibitem [{\citenamefont {Sanders}\ \emph {et~al.}(2007)\citenamefont
  {Sanders}, \citenamefont {Verhulst},\ and\ \citenamefont
  {Murdock}}]{sanders2007averaging}%
  \BibitemOpen
  \bibfield  {author} {\bibinfo {author} {\bibfnamefont {J.~A.}\ \bibnamefont
  {Sanders}}, \bibinfo {author} {\bibfnamefont {F.}~\bibnamefont {Verhulst}}, \
  and\ \bibinfo {author} {\bibfnamefont {J.}~\bibnamefont {Murdock}},\
  }\href@noop {} {\emph {\bibinfo {title} {Averaging methods in nonlinear
  dynamical systems}}},\ Vol.~\bibinfo {volume} {59}\ (\bibinfo  {publisher}
  {Springer},\ \bibinfo {year} {2007})\BibitemShut {NoStop}%
\bibitem [{\citenamefont {Pavliotis}\ and\ \citenamefont
  {Stuart}(2008)}]{pavliotis2008multiscale}%
  \BibitemOpen
  \bibfield  {author} {\bibinfo {author} {\bibfnamefont {G.}~\bibnamefont
  {Pavliotis}}\ and\ \bibinfo {author} {\bibfnamefont {A.}~\bibnamefont
  {Stuart}},\ }\href@noop {} {\emph {\bibinfo {title} {Multiscale methods:
  averaging and homogenization}}}\ (\bibinfo  {publisher} {Springer},\ \bibinfo
  {year} {2008})\BibitemShut {NoStop}%
\bibitem [{\citenamefont {Werbos}(1990)}]{werbos1990backpropagation}%
  \BibitemOpen
  \bibfield  {author} {\bibinfo {author} {\bibfnamefont {P.~J.}\ \bibnamefont
  {Werbos}},\ }\href@noop {} {\bibfield  {journal} {\bibinfo  {journal} {Proc.
  IEEE}\ }\textbf {\bibinfo {volume} {78}},\ \bibinfo {pages} {1550} (\bibinfo
  {year} {1990})}\BibitemShut {NoStop}%
\bibitem [{\citenamefont {Larger}\ \emph {et~al.}(2017)\citenamefont {Larger},
  \citenamefont {Bayl{\'o}n-Fuentes}, \citenamefont {Martinenghi},
  \citenamefont {Udaltsov}, \citenamefont {Chembo},\ and\ \citenamefont
  {Jacquot}}]{larger2017high}%
  \BibitemOpen
  \bibfield  {author} {\bibinfo {author} {\bibfnamefont {L.}~\bibnamefont
  {Larger}}, \bibinfo {author} {\bibfnamefont {A.}~\bibnamefont
  {Bayl{\'o}n-Fuentes}}, \bibinfo {author} {\bibfnamefont {R.}~\bibnamefont
  {Martinenghi}}, \bibinfo {author} {\bibfnamefont {V.~S.}\ \bibnamefont
  {Udaltsov}}, \bibinfo {author} {\bibfnamefont {Y.~K.}\ \bibnamefont
  {Chembo}}, \ and\ \bibinfo {author} {\bibfnamefont {M.}~\bibnamefont
  {Jacquot}},\ }\href@noop {} {\bibfield  {journal} {\bibinfo  {journal}
  {Physical Review X}\ }\textbf {\bibinfo {volume} {7}},\ \bibinfo {pages}
  {011015} (\bibinfo {year} {2017})}\BibitemShut {NoStop}%
\bibitem [{\citenamefont {Tanaka}\ \emph {et~al.}(2019)\citenamefont {Tanaka},
  \citenamefont {Yamane}, \citenamefont {H{\'e}roux}, \citenamefont {Nakane},
  \citenamefont {Kanazawa}, \citenamefont {Takeda}, \citenamefont {Numata},
  \citenamefont {Nakano},\ and\ \citenamefont {Hirose}}]{tanaka2019recent}%
  \BibitemOpen
  \bibfield  {author} {\bibinfo {author} {\bibfnamefont {G.}~\bibnamefont
  {Tanaka}}, \bibinfo {author} {\bibfnamefont {T.}~\bibnamefont {Yamane}},
  \bibinfo {author} {\bibfnamefont {J.~B.}\ \bibnamefont {H{\'e}roux}},
  \bibinfo {author} {\bibfnamefont {R.}~\bibnamefont {Nakane}}, \bibinfo
  {author} {\bibfnamefont {N.}~\bibnamefont {Kanazawa}}, \bibinfo {author}
  {\bibfnamefont {S.}~\bibnamefont {Takeda}}, \bibinfo {author} {\bibfnamefont
  {H.}~\bibnamefont {Numata}}, \bibinfo {author} {\bibfnamefont
  {D.}~\bibnamefont {Nakano}}, \ and\ \bibinfo {author} {\bibfnamefont
  {A.}~\bibnamefont {Hirose}},\ }\href@noop {} {\bibfield  {journal} {\bibinfo
  {journal} {Neural Net.}\ }\textbf {\bibinfo {volume} {115}},\ \bibinfo
  {pages} {100 } (\bibinfo {year} {2019})}\BibitemShut {NoStop}%
\bibitem [{\citenamefont {Pikovsky}\ \emph {et~al.}(2003)\citenamefont
  {Pikovsky}, \citenamefont {Kurths},\ and\ \citenamefont
  {Rosenblum}}]{pikovsky2003synchronization}%
  \BibitemOpen
  \bibfield  {author} {\bibinfo {author} {\bibfnamefont {A.}~\bibnamefont
  {Pikovsky}}, \bibinfo {author} {\bibfnamefont {J.}~\bibnamefont {Kurths}}, \
  and\ \bibinfo {author} {\bibfnamefont {M.}~\bibnamefont {Rosenblum}},\
  }\href@noop {} {\emph {\bibinfo {title} {Synchronization: a universal concept
  in nonlinear sciences}}},\ Vol.~\bibinfo {volume} {12}\ (\bibinfo
  {publisher} {Cambridge university press},\ \bibinfo {year}
  {2003})\BibitemShut {NoStop}%
\bibitem [{\citenamefont {Rivkind}\ and\ \citenamefont
  {Barak}(2017)}]{rivkind2017local}%
  \BibitemOpen
  \bibfield  {author} {\bibinfo {author} {\bibfnamefont {A.}~\bibnamefont
  {Rivkind}}\ and\ \bibinfo {author} {\bibfnamefont {O.}~\bibnamefont
  {Barak}},\ }\href@noop {} {\bibfield  {journal} {\bibinfo  {journal} {Phys.
  Rev. Lett.}\ }\textbf {\bibinfo {volume} {118}},\ \bibinfo {pages} {258101}
  (\bibinfo {year} {2017})}\BibitemShut {NoStop}%
\bibitem [{\citenamefont {Jiang}\ and\ \citenamefont
  {Lai}(2019)}]{jiang2019model}%
  \BibitemOpen
  \bibfield  {author} {\bibinfo {author} {\bibfnamefont {J.}~\bibnamefont
  {Jiang}}\ and\ \bibinfo {author} {\bibfnamefont {Y.-C.}\ \bibnamefont
  {Lai}},\ }\href@noop {} {\bibfield  {journal} {\bibinfo  {journal} {Phys.
  Rev. Res.}\ }\textbf {\bibinfo {volume} {1}},\ \bibinfo {pages} {033056}
  (\bibinfo {year} {2019})}\BibitemShut {NoStop}%
\bibitem [{Note1()}]{Note1}%
  \BibitemOpen
  \bibinfo {note} {In \cite {ottLyapunov,ottAttractor} it is suggested that
  this choice respects the symmetries of the Lorenz system. In general, any reasonably nonlinear function of $r_i$ should suffice \cite
  {jaeger2001echo}}\BibitemShut {NoStop}%
\bibitem [{\citenamefont {Milano}\ and\ \citenamefont
  {Koumoutsakos}(2002)}]{milano2002neural}%
  \BibitemOpen
  \bibfield  {author} {\bibinfo {author} {\bibfnamefont {M.}~\bibnamefont
  {Milano}}\ and\ \bibinfo {author} {\bibfnamefont {P.}~\bibnamefont
  {Koumoutsakos}},\ }\href@noop {} {\bibfield  {journal} {\bibinfo  {journal}
  {J. Comput. Phys.}\ }\textbf {\bibinfo {volume} {182}},\ \bibinfo {pages} {1}
  (\bibinfo {year} {2002})}\BibitemShut {NoStop}%
\bibitem [{\citenamefont {Wan}\ \emph {et~al.}(2018)\citenamefont {Wan},
  \citenamefont {Vlachas}, \citenamefont {Koumoutsakos},\ and\ \citenamefont
  {Sapsis}}]{wan2018data}%
  \BibitemOpen
  \bibfield  {author} {\bibinfo {author} {\bibfnamefont {Z.~Y.}\ \bibnamefont
  {Wan}}, \bibinfo {author} {\bibfnamefont {P.}~\bibnamefont {Vlachas}},
  \bibinfo {author} {\bibfnamefont {P.}~\bibnamefont {Koumoutsakos}}, \ and\
  \bibinfo {author} {\bibfnamefont {T.}~\bibnamefont {Sapsis}},\ }\href@noop {}
  {\bibfield  {journal} {\bibinfo  {journal} {PloS one}\ }\textbf {\bibinfo
  {volume} {13}},\ \bibinfo {pages} {e0197704} (\bibinfo {year}
  {2018})}\BibitemShut {NoStop}%
\bibitem [{\citenamefont {Weymouth}\ and\ \citenamefont
  {Yue}(2013)}]{weymouth2013physics}%
  \BibitemOpen
  \bibfield  {author} {\bibinfo {author} {\bibfnamefont {G.~D.}\ \bibnamefont
  {Weymouth}}\ and\ \bibinfo {author} {\bibfnamefont {D.~K.~P.}\ \bibnamefont
  {Yue}},\ }\href@noop {} {\bibfield  {journal} {\bibinfo  {journal} {J. Ship
  Res.}\ }\textbf {\bibinfo {volume} {57}},\ \bibinfo {pages} {1} (\bibinfo
  {year} {2013})}\BibitemShut {NoStop}%
\bibitem [{\citenamefont {Boffetta}\ \emph {et~al.}(1998)\citenamefont
  {Boffetta}, \citenamefont {Giuliani}, \citenamefont {Paladin},\ and\
  \citenamefont {Vulpiani}}]{boffetta1998extension}%
  \BibitemOpen
  \bibfield  {author} {\bibinfo {author} {\bibfnamefont {G.}~\bibnamefont
  {Boffetta}}, \bibinfo {author} {\bibfnamefont {P.}~\bibnamefont {Giuliani}},
  \bibinfo {author} {\bibfnamefont {G.}~\bibnamefont {Paladin}}, \ and\
  \bibinfo {author} {\bibfnamefont {A.}~\bibnamefont {Vulpiani}},\ }\href@noop
  {} {\bibfield  {journal} {\bibinfo  {journal} {J. Atmos. Sci.}\ }\textbf
  {\bibinfo {volume} {55}},\ \bibinfo {pages} {3409} (\bibinfo {year}
  {1998})}\BibitemShut {NoStop}%
\bibitem [{\citenamefont {Carlu}\ \emph {et~al.}(2019)\citenamefont {Carlu},
  \citenamefont {Ginelli}, \citenamefont {Lucarini},\ and\ \citenamefont
  {Politi}}]{ginelli}%
  \BibitemOpen
  \bibfield  {author} {\bibinfo {author} {\bibfnamefont {M.}~\bibnamefont
  {Carlu}}, \bibinfo {author} {\bibfnamefont {F.}~\bibnamefont {Ginelli}},
  \bibinfo {author} {\bibfnamefont {V.}~\bibnamefont {Lucarini}}, \ and\
  \bibinfo {author} {\bibfnamefont {A.}~\bibnamefont {Politi}},\ }\href
  {\doibase 10.5194/npg-26-73-2019} {\bibfield  {journal} {\bibinfo  {journal}
  {Nonlin. Proc. Geophys.}\ }\textbf {\bibinfo {volume} {26}},\ \bibinfo
  {pages} {73} (\bibinfo {year} {2019})}\BibitemShut {NoStop}%
\bibitem [{\citenamefont {Carmichael}\ \emph {et~al.}(2019)\citenamefont
  {Carmichael}, \citenamefont {Syed},\ and\ \citenamefont
  {Kudithipudi}}]{carmichael2019analysis}%
  \BibitemOpen
  \bibfield  {author} {\bibinfo {author} {\bibfnamefont {Z.}~\bibnamefont
  {Carmichael}}, \bibinfo {author} {\bibfnamefont {H.}~\bibnamefont {Syed}}, \
  and\ \bibinfo {author} {\bibfnamefont {D.}~\bibnamefont {Kudithipudi}},\ }in\
  \href@noop {} {\emph {\bibinfo {booktitle} {Proc. 7th Annual Neuro-inspired
  Comput. Elements Workshop}}}\ (\bibinfo {year} {2019})\ pp.\ \bibinfo {pages}
  {1--10}\BibitemShut {NoStop}%
\bibitem [{\citenamefont {Lu}\ \emph {et~al.}(2017)\citenamefont {Lu},
  \citenamefont {Pathak}, \citenamefont {Hunt}, \citenamefont {Girvan},
  \citenamefont {Brockett},\ and\ \citenamefont {Ott}}]{lu2017reservoir}%
  \BibitemOpen
  \bibfield  {author} {\bibinfo {author} {\bibfnamefont {Z.}~\bibnamefont
  {Lu}}, \bibinfo {author} {\bibfnamefont {J.}~\bibnamefont {Pathak}}, \bibinfo
  {author} {\bibfnamefont {B.}~\bibnamefont {Hunt}}, \bibinfo {author}
  {\bibfnamefont {M.}~\bibnamefont {Girvan}}, \bibinfo {author} {\bibfnamefont
  {R.}~\bibnamefont {Brockett}}, \ and\ \bibinfo {author} {\bibfnamefont
  {E.}~\bibnamefont {Ott}},\ }\href@noop {} {\bibfield  {journal} {\bibinfo
  {journal} {Chaos}\ }\textbf {\bibinfo {volume} {27}},\ \bibinfo {pages}
  {041102} (\bibinfo {year} {2017})}\BibitemShut {NoStop}%
\bibitem [{\citenamefont {Meneveau}\ and\ \citenamefont
  {Katz}(2000)}]{meneveau2000scale}%
  \BibitemOpen
  \bibfield  {author} {\bibinfo {author} {\bibfnamefont {C.}~\bibnamefont
  {Meneveau}}\ and\ \bibinfo {author} {\bibfnamefont {J.}~\bibnamefont
  {Katz}},\ }\href@noop {} {\bibfield  {journal} {\bibinfo  {journal} {Annu.
  Rev. Fluid Mech.}\ }\textbf {\bibinfo {volume} {32}},\ \bibinfo {pages} {1}
  (\bibinfo {year} {2000})}\BibitemShut {NoStop}%
\bibitem [{\citenamefont {Sagaut}(2006)}]{sagaut2006large}%
  \BibitemOpen
  \bibfield  {author} {\bibinfo {author} {\bibfnamefont {P.}~\bibnamefont
  {Sagaut}},\ }\href@noop {} {\emph {\bibinfo {title} {Large eddy simulation
  for incompressible flows: an introduction}}}\ (\bibinfo  {publisher}
  {Springer},\ \bibinfo {year} {2006})\BibitemShut {NoStop}%
\bibitem [{\citenamefont {Biferale}\ \emph {et~al.}(2017)\citenamefont
  {Biferale}, \citenamefont {Mailybaev},\ and\ \citenamefont
  {Parisi}}]{biferale2017optimal}%
  \BibitemOpen
  \bibfield  {author} {\bibinfo {author} {\bibfnamefont {L.}~\bibnamefont
  {Biferale}}, \bibinfo {author} {\bibfnamefont {A.~A.}\ \bibnamefont
  {Mailybaev}}, \ and\ \bibinfo {author} {\bibfnamefont {G.}~\bibnamefont
  {Parisi}},\ }\href@noop {} {\bibfield  {journal} {\bibinfo  {journal} {Phys.
  Rev. E}\ }\textbf {\bibinfo {volume} {95}},\ \bibinfo {pages} {043108}
  (\bibinfo {year} {2017})}\BibitemShut {NoStop}%
\bibitem [{Note2()}]{Note2}%
  \BibitemOpen
  \bibinfo {note} {Notice that if $T_t/\Delta t<D_R$ at least one eigenvalue is
  zero.}\BibitemShut {Stop}%
\bibitem [{\citenamefont {Kennedy}\ and\ \citenamefont
  {Eberhart}(1995)}]{kennedy1995particle}%
  \BibitemOpen
  \bibfield  {author} {\bibinfo {author} {\bibfnamefont {J.}~\bibnamefont
  {Kennedy}}\ and\ \bibinfo {author} {\bibfnamefont {R.}~\bibnamefont
  {Eberhart}},\ }in\ \href@noop {} {\emph {\bibinfo {booktitle} {Proceedings of
  ICNN'95-International Conference on Neural Networks}}},\ Vol.~\bibinfo
  {volume} {4}\ (\bibinfo {organization} {IEEE},\ \bibinfo {year} {1995})\ pp.\
  \bibinfo {pages} {1942--1948}\BibitemShut {NoStop}%
\end{thebibliography}
\end{document}